\documentclass[10pt,conference]{IEEEtran}
\IEEEoverridecommandlockouts
\usepackage{cite}
\usepackage{amsmath,amssymb,amsfonts}
\usepackage{algorithmic}
\usepackage{graphicx}
\usepackage{textcomp}
\usepackage{xcolor}
\usepackage{balance}
\def\BibTeX{{\rm B\kern-.05em{\sc i\kern-.025em b}\kern-.08em
    T\kern-.1667em\lower.7ex\hbox{E}\kern-.125emX}}

\newcommand{\mytitle}{Input Debugging via Rich %and Fast
Failure Feedback}
\renewcommand{\mytitle}{Repairing Inputs}
\renewcommand{\mytitle}{Repairing Inputs with Input Synthesis}
\renewcommand{\mytitle}{Repairing Inputs via Synthesis}
\renewcommand{\mytitle}{Input Repair via Synthesis and Lightweight \\ Error Feedback}

\usepackage{listings}
\usepackage{xspace}
\usepackage{array,multirow,graphicx}
\usepackage{float}
\usepackage{framed}
\usepackage{tikz,pgfplots,pgfplotstable}
\usetikzlibrary{calc,positioning,patterns}
\usepackage{hyperref}
\usepackage{cleveref}
\usepackage{boxedminipage}
\usepackage[inline]{enumitem}
\newenvironment{result}{\begin{framed}\centering\it}{\end{framed}}

\def\|#1|{\textit{#1}}
\def\<#1>{\texttt{#1}}

\newcounter{todocounter}
\newcommand{\todo}[1]{\marginpar{$|$}\textcolor{red}{\stepcounter{todocounter}\footnote[\thetodocounter]{\textcolor{red}{\textbf{TODO }}\textit{#1}}}}
\newcommand{\done}[1]{\marginpar{$*$}\textcolor{green}{\stepcounter{todocounter}\footnote[\thetodocounter]{\textcolor{black}{\textbf{DONE }}\textit{#1}}}}

\newcommand{\revise}[1]{#1}

\renewcommand{\done}[1]{} % comment to see responses.

\IfFileExists{SUBMIT}{
\renewcommand{\todo}[1]{}
\renewcommand{\done}[1]{}
}{
}

\newcommand{\ddmin}{\textit{ddmin}\xspace}
\newcommand{\test}{\textit{test}\xspace}

\usepackage{pifont}
\newcommand{\pass}{\text{\ding{52}}\xspace}
\newcommand{\fail}{\text{\ding{56}}\xspace}
\newcommand{\unresolved}{\lower0.1ex\hbox{\includegraphics*[height=1.7ex]{question.pdf}}}

\newcommand{\cpass}{{c_{\scriptscriptstyle \pass}}}
\newcommand{\cfail}{{c_{\scriptscriptstyle \fail}}}
\newcommand{\dpass}{{c'_{\scriptscriptstyle \pass}}}

\newcommand{\approach}{\textsc{FSynth}\xspace}

\def\ddmin{DDMin\xspace}
\newcommand{\ddmax}{\textit{DDMax}\xspace}
\newcommand{\ddmaxg}{\textit{DDmaxG}\xspace}

\newcommand{\brepair}{\textsc{FSynth}\xspace}

\tikzset{inlinenode/.style={draw=white,text=black,fill=light-gray,inner sep=.1em,outer sep=0em}}

\tikzset{%
simpletext/.style={draw=none,text=black,font=\normalfont\normalsize,align=center},
gparsetreenode/.style={minimum width=6mm,minimum height=5mm},
lparsetreenode/.style={gparsetreenode,simpletext,rectangle,draw=white,fill=white,align=center},
lparsetreeerrornode/.style={lparsetreenode,font=\bfseries},
lparsetreephantomnode/.style={gparsetreenode,edge from parent/.append style={draw=none},shape=coordinate,minimum width=15mm},
lparsetreestrikethrough/.style={draw=black,thick},
lparsetreedeletednode/.style={lparsetreenode,append after command={\pgfextra \draw[lparsetreestrikethrough] (\tikzlastnode.north west) -- (\tikzlastnode.south east); \draw[lparsetreestrikethrough] (\tikzlastnode.north east) -- (\tikzlastnode.south west);\endpgfextra}},
lparsetree/.style={node distance=5mm,level distance=10mm,every node/.style={lparsetreenode},edge from parent/.style={draw=black,-latex,shorten >=.5mm}
},
lflowchartnode/.style={lparsetreenode,draw=black,rounded corners=.5pt},
blockdiagramlines/.style={draw,stroke=black,line width=1.2pt},
blockdiagramarrow/.style={blockdiagramlines,->},
blockdiagramdashedarrow/.style={blockdiagramarrow,dashed},
blockdiagramannot/.style={blockdiagramlines,text=black,align=center},
blockdiagramblock/.style={lflowchartnode,blockdiagramannot,minimum width=1.5cm,minimum height=0.5cm,text width=1.9cm},
blockdiagrammicroblock/.style={blockdiagramannot,font=\tiny,minimum width=1.5cm,minimum height=.5cm,rounded corners=.5pt},
blockdiagramarrowcaption/.style={font=\scriptsize\sffamily,inner sep=1.5pt,text=black},
blockdiagramouterbox/.style={blockdiagramlines,densely dotted,line width=.7pt},
blockdiagramouterboxcaption/.style={blockdiagramarrowcaption,font=\itshape\scriptsize,inner sep=1pt},
pics/numbering/.style args={#1}{code={
    \node[draw=black,shape=circle,fill=white,text=black,font=\ttfamily\scriptsize,inner sep=1pt,text width=8pt,align=center,outer sep=0] (-number) {#1};
}}}

\definecolor{light-gray}{gray}{0.87}
\makeatletter
\newcommand\letterboxed[1]{%
\setlength{\fboxsep}{0pt}%
  \@tfor\@ii:=#1\do{%
    \fcolorbox{white}{light-gray}{\texttt{\strut\@ii}}%
  }%
}

\newcommand\letterboxedinTable[1]{%
\setlength{\fboxsep}{0pt}%
  \@tfor\@ii:=#1\do{%
    \fcolorbox{white}{light-gray}{\tiny \texttt{\strut\@ii}}%
  }%
}

\makeatother

\usepackage[utf8]{inputenc}

\begin{document}

\title{\mytitle}
\author{\IEEEauthorblockN{Lukas Kirschner}
\IEEEauthorblockA{\textit{Saarland University}, Germany\\
s8lukirs@stud.uni-saarland.de}
\\
\IEEEauthorblockN{Ezekiel Soremekun}
\IEEEauthorblockA{\textit{University of Luxembourg}, Luxembourg\\
ezekiel.soremekun@uni.lu}
\and
\IEEEauthorblockN{Rahul Gopinath}
\IEEEauthorblockA{\textit{University of Sydney}, Australia\\
rahul.gopinath@sydney.edu.au}
\\
\IEEEauthorblockN{Andreas Zeller}
\IEEEauthorblockA{\textit{CISPA Helmholtz Center for Information Security}, Germany\\
zeller@cispa.de}
}
\maketitle

\begin{abstract}
\done{For the title, I'd definitely include ``synthesis'', as in ``Repairing Inputs with Input Synthesis'' -- AZ}
%Data cleansing and data 
%in fields that aggregate data from multiple resources (e.g., data warehousing). 
Oftentimes,
input data may ostensibly conform to a given input format, but cannot be parsed by a conforming program, for instance, due to human error or data corruption.
In such cases, a data engineer \revise{is tasked with \textit{input repair}, i.e., she has to manually repair the corrupt data such that
it follows a given format, and hence can be processed by the conforming program.}
Such manual repair can be time-consuming and
error-prone. 
%When a program input is invalid, developers are tasked with repairing such inputs.  
\revise{In particular, input repair is challenging without an input specification (e.g., input grammar) or program analysis}.

In this work, we show that incorporating lightweight failure feedback (e.g., input incompleteness) % \todo{what is lightweight error-feedback?} 
to parsers is sufficient to repair any corrupt input data with maximal closeness to the
semantics of the input data. We propose an approach (called \brepair) that leverages \emph{lightweight error-feedback} and \emph{input synthesis} to repair invalid inputs. 
\revise{\brepair is \textit{grammar-agnostic} and it does not require program analysis}. 
%\todo{We need to clarify the feedback requirement, correctness and incompleteness}
Given a conforming program, %\todo{it is not clear, what is a conforming program, a parser or any program conforming to the input format??}, 
and any invalid input, \brepair provides a set
of repairs prioritized by the distance of the repair from the original input.

We evaluate \brepair on 806 (real-world) invalid inputs using four well-known input formats, namely INI, TinyC, SExp, and cJSON. In our evaluation, we found that \approach 
\revise{recovers 91\% of valid input data.} %and it 
\approach is also highly effective and efficient in input repair: It repairs 77\% of invalid inputs %and recovers 91\% of input data, 
within four minutes. It is up to 35\% more effective than \ddmax, the previously best-known approach.
Overall, our approach addresses several
limitations of
\ddmax,
both in terms of what it can repair, as well as
in terms of the set of repairs offered.
\end{abstract}

\section{Introduction}
\label{sec:intro}

\revise{
Input data is prone to unintended errors. Such errors may be introduced when
the data is being created (by humans or by buggy programs), modified (by external actors)
or transmitted (via flawed networks)~\cite{scaffidi2008accommodating}. 
For instance, several inputs are created by hand, leading to \textit{invalid inputs}~\cite{mucslu2015preventing}. Invalid input data may also be caused by  disagreements among data sources on the format specification which leads to different %variances in the
implementations of the specification. For example, JSON libraries implement slightly different
definitions of the JSON formats~\cite{harrand2021behavioral,seriot2016parsing}, different database systems support %, each supporting a
slightly different SQL formats~\cite{arvin2018comparison}, and  various
\<C> compilers provide slightly different interpretation of the \<C> language.
These problems lead to invalid inputs that cannot be processed by their conforming programs or consumed by end-users (e.g., developers). 
%subsequently, developers and software can not consume such inputs. 
}

\revise{
Given such invalid inputs that are \emph{almost} but \emph{not quite}
parsable, developers are saddled with the task of \textit{input repair}. Input
repair is particularly important due to the high prevalence of invalid inputs
in software practice~\cite{kirschner2020debugging}. It can be challenging
to recover the valid portion of invalid inputs
automatically~\cite{scaffidi2008topes}, and developers often have to
\emph{manually} repair such inputs~\cite{kirschner2020debugging}, which is
time-consuming and error-prone.
}
%\revise{
%Specifically, t
%To repair invalid inputs, %consuming such resources %, and aggregating them 
%often involves trying to
%developers \textit{manually} try to 
%maximally parse them by minimizing parse errors~\cite{kirschner2020debugging}, this manual process is time-consuming and error-prone.}

\revise{
%On one hand, g
Given a formal grammar for inputs, grammar-based input repair approaches such as 
\emph{error-correcting}
parsers~\cite{aho1972minimum,diekmann2020dont,parr2011ll}, can
repair invalid inputs.
%parser that conforms to an available 
%a formal specification such as a context-free
%grammar. 
%, and the grammar is available.
} 
%context-free grammar, and an acceptable formal grammar is available,
%the challenge of repairing the input and making it parsable can be solved using
%.
The main idea of such parsers is to generate a universal grammar that captures
any mutation of the base grammar. Any parse of the input string that employs a
mutation is penalized, and the parse with minimal mutation is chosen as the
best parse, and the corresponding mutation the best repair. The \textit{limitation} of
this approach is that\textit{ it assumes the existence of a formal grammar} in the first
place, which completely captures the intended structure.
%However, %\recheck{
%formats such as markdown~\cite{gruber2004markdown} does not have
%an official formal grammar;
%%}\todo{how do you mean? this is not true, there are markdown grammars, ANTLR also has one, Nikolas used some in his Tribble eval too.};
%documents in languages such as \<C> can contain semantic
However, %the formal grammar
\textit{this assumption may not hold in practice} due to several reasons.
Firstly, %or stance,
certain input formats (e.g., URL standard from WHATWG~\cite{whatwgurl}) may not have
an official formal grammar specification.
%Others may have numerous standards
%(e.g., markdown~\cite{gruber2004markdown}), any of which may be followed by the
%artifact in question.
%Documents in languages (such as \<C>) can contain semantic
%information that is not captured in the context-free grammar, and may not be
%discardable. Some input constraints (e.g., the number of columns in a \<CSV> file) may
%not be captured in a grammar. Finally, e
Even when the base grammar is context-free
%(e.g., JSON)
 the serializer and deserializer may implement common
subsets beyond the base grammar (e.g., comments and unquoted keys in JSON) which may
contain important information. 
\revise{
Hence, grammar-based 
%these 
input repair approaches %relying on error-correcting parsers may
%be 
are suboptimal for fixing invalid inputs in practice.}
% In this work, we address this challenge via a \textit{grammar-agnostic} input repair method.
% (like \ddmax ~\cite{kirschner2020debugging}). 
%that addresses this limitation, among many others.
%}
%This may be because the constraints in
%the input may not be fully captured in the grammar as in the case of CSV where
%the number of columns may be fixed, or because such a grammar is unavailable
%(as in the case of markdown)

\revise{
%On the other hand, l
Black-box and language-agnostic input repair methods (e.g., \textit{lexical} \ddmax~\cite{kirschner2020debugging}) repair invalid inputs without a base grammar or program analysis.}
%In circumstances where one cannot rely on a base grammar, the only option so far
%has been \textit{lexical} \ddmax~\cite{kirschner2020debugging}. 
\ddmax works similar to
Delta Debugging~\cite{zeller2002simplifying}, but in reverse. The idea
is that given a corrupt input which induces a parse error, and a way to
decompose the input into independent fragments (called deltas ($\delta$)), one can
successively minimize the parse-error inducing part of the input resulting in a
maximal parsing input.
%TODO Use textcite here:
%More importantly, Kirschner et al.~\cite{kirschner2020debugging} and Scaffidi et al.~\cite{scaffidi2008topes} demonstrate the prevalence of invalid inputs in the wild and the importance of input repair in practice. % without grammars or program analysis.
%using \ddmax. %}.
Although \ddmax works very well for most inputs, %, it could repair 69\% of input files and recovered about 78\% of input data.
%However,
we observe it %that \ddmax
has certain limitations which inhibit it from completely repairing any invalid inputs. % in practice.
\Cref{tab:ddmaxlimitations} and \Cref{sec:rich-input-structures} illustrate %some of
these limitations with simplified invalid inputs. % as examples.
Notably, two major \ddmax limitations include its (1) \textit{limited repair operation} (only deletion), and (2) inability to completely repair \textit{rich input structures} (e.g., multiple faults) due to inherent assumptions. 
Our approach addresses these limitations of \ddmax, which we now discuss in detail.
% (e.g., its search span)..
%\textit{multiple faults in input data}, and (3)
%\textit{inherent assumptions about the richness of input structures} (e.g., its search span).

\begin{table*}\centering

\caption{\ddmax vs. \approach: examples showing limitations of \ddmax and the strengths of \approach}
{\scriptsize  %footnotesize
\begin{tabular}{|l | c | c | l |}
\hline
\textbf{Example} & \textbf{\ddmax repair} & \textbf{\brepair repair} & \textbf{\ddmax limitation} \\
\hline
\letterboxedinTable{\{\ "name":\ "Dave"\ "age":\ 42\ \}} &
%\letterboxed{\{\ "name":\ "Daveage":\ 42\ \}} &
\letterboxedinTable{\ \ \ \ 42\ }  &
\letterboxedinTable{\{\ "name":\ "Dave"\ ,"age":\ 42\ \}} &
Limited repair options (deletion) \\
%\hline
\letterboxedinTable{\{\ "item":\ "Apple",\ "price":\ ***3.45\}} & \letterboxedinTable{\ \ \ \
3.45\ } & \letterboxedinTable{\{\ "item":\ "Apple",\ "price":\ 3.45\}} & Rich structure
(spans) \\
\letterboxedinTable{\{"ABCD":[*"1,2,3,4,5,6"]*\}} &
\letterboxedinTable{123456} &
\letterboxedinTable{\{"ABCD":["1,2,3,4,5,6"]\}} &
Rich Structure (multiple-faults) \\

\hline
\end{tabular}}
\label{tab:ddmaxlimitations}
%\vspace{-0.5cm}
\end{table*}

Firstly, \ddmax's repair is restricted to deletion, which makes it sub-optimal for repairing some invalid inputs. 
%The problem is that 
While \ddmax can repair invalid inputs due to 
%in many circumstances such as
%disk corruption, some input characters are more likely to be \emph{changed} or \emph{deleted}
%rather than \emph{inserted.} Hence, by handling only 
spurious insertions, it %\ddmax 
%misses an important repair requirement - for 
fails to repair input 
invalidity that is due to \emph{changed} or \emph{deleted} fragments. Consider the sample invalid input in \textit{row one} of \Cref{tab:ddmaxlimitations} (i.e., \letterboxed{\{\ "name":\ "Dave"\ "age":\ 42\ \}}) which is invalid because of a \textit{missing comma separator} between the two values in the JSON object. Due to the limited repair options of \ddmax (i.e., only deletion), its resulting repair (i.e., \letterboxed{\ \ \ \ 42\ }) is \textit{sub-optimal} leading to a huge (75\%) data loss (21 out of 28 bytes). %, % $\approx$ .
%  data loss).
% , only seven out of 28 bytes recovered).
\revise{
This work addresses this challenge % with an additional repair requirement to 
by supporting the synthesis of missing input fragments. 
}

Secondly,
%As example,
%a major limitation is that
\ddmax is modeled after \ddmin, where one of
the unstated assumptions is that the $\delta$ fragments that are not
contributing to the failure observed can be independently removed without
affecting the failure observed. When this assumption is not met, (i.e. where
the inputs have a rich structure) the minimal fragment produced by \ddmin
can be suboptimal. %Hence, the h
%Hierarchical delta debugging (HDD)~\cite{misherghi2006hdd}
%and its descendants address this concern for reducing failure-inducing inputs, however, this problem remains open for input repair. 
Similarly, 
%a similar constraint also applies to \ddmax, where
%the unstated assumption is that 
\textit{\ddmax assumes each non-failure-inducing fragment
can be added to the passing subset without inducing a failure}. When this
assumption is not met (as in the case of inputs with a rich structure such as
conforming to a grammar) the repairs produced can be suboptimal as the
examples show.
%\recheck{
As an example, consider the invalid input in \textit{row three} of \Cref{tab:ddmaxlimitations} (i.e.,  \letterboxed{\{"ABCD":[*"1,2,3,4,5,6"]*\}}). Due to the \textit{multiple faults (two \letterboxed{*}s)} in this input, the resulting \ddmax repair (i.e., \letterboxed{123456}) is \textit{sub-optimal} causing a 77\% data loss (20 out of 26 bytes).
%\revise{
%This paper addresses this limitation to recover almost all valid data. 
% via a combination of lightweight error-feedback and input synthesis. 
%}

%&
% &
%\letterboxed{\{"ABCD":["1,2,3,4,5,6"]\}} &
%
%...\todo{we need to decide the most important limitation to buttress with an example here ... I think one that requires insertion and incompleteness error-feedback is the most important differentiator, or a hybrid example that also contains multiple faults ... } \todo{it should also be the same example to motivate and illustrate the way \brepair works (after the par. )}
% we address these limitations ...
%}

%\todo{to Move/restructure}
%However, the existence of \ddmax suggests that repairing inputs without a formal
%grammar at hand is an important task. Unfortunately, most documents in the real
%world have some syntactic or semantic constraints that makes satisfying the
%implied assumption of \ddmax (independence of fragments) unsatisfiable.
%The second limitation of \ddmax -- the repair is restricted to deletion -- makes it
%even more suboptimal to use. The problem is that in many circumstances such as
%disk corruption, some input characters are more likely to be changed or deleted
%rather than inserted. Hence by handling only spurious insertions, \ddmax misses
%an important repair requirement.

This paper introduces the \brepair approach\footnote{
%\recheck{
\brepair denotes ``\underline{\textbf{\textsc{F}}}eedback-driven Input \underline{\textbf{\textsc{SYNTH}}}esis''}
%} 
%Fixing invalid input via input SYNTHesis".}
% Feedback driven SYNThesis of repairs
%}
%\todo{remember to say the meaning or derivation of the name \approach}
to address the %provide a solution that addresses these
limitations of \ddmax via \textit{lightweight error-feedback and input synthesis}. The key insight of this approach is to complete \textit{semantic} repair of invalid inputs using lightweight failure feedback such as the \textit{validity, incompleteness, and incorrectness} checks of input fragments. Given an invalid input and a conforming program, \approach performs test experiments of input subsets and candidate insertions to provide a set of repairs prioritized by the distance of the repair from the original input.
Unlike \ddmax, \brepair does not have the \textit{implied assumption of independence of
fragments}. Hence, it does not face any limitation when given invalid inputs
that should conform to a rich structure, and %. It also
does not stumble when faced
with multiple faults. Finally, it %\brepair
provides both deletion and
synthesis as repair options making it an optimal algorithm in the toolbox of
data engineers.
%\todo{what else is the main ingredient or key insight of the approach}

%\todo{to Move/restructure}
%This paper %We
%shows that corrupted inputs with rich structure can be repaired adequately if
%the input processor can provide at least some indication of progress.
%That is, as with \ddmax, we require the input processor to indicate if the input
%is valid.  However, when given an invalid input, we require the input processor
%to indicate whether the input is merely \emph{incomplete}---that is, the input
%is a prefix of a valid input---or is \emph{incorrect}---that is, no suffix to
%this input will result in a valid input.
%
%This requirement is not hard to accomplish. In most cases, parsers already
%provide precise information as to the point at which the parse failed, which
%can be used directly in our algorithm. In the cases where parers do not provide
%this information, obtaining this information by external instrumentation is
%possible as demonstrated by Bj\"orn et al.~\cite{mathis2019parser}. If
%the program is hard to instrument, adding failure feedback that we require is
%not difficult~\todo{
%The arxiv link is~\cite{gopinath2020fuzzing} but we can't cite it because
%bFuzzer is getting submitted at FSE too. -> alright, we can cite it after if accepted}.

%\recheck{
For instance, consider the invalid input in \textit{row one} of \Cref{tab:ddmaxlimitations} (i.e., \letterboxed{\{\ "name":\ "Dave"\ "age":\ 42\ \}}) which is invalid because of a \textit{missing comma separator} between the two values in the JSON object. While \ddmax could not completely repair this input, \brepair could complete a repair via input synthesis. % owing to its input synthesis component.
%\todo{to Move/restructure}
%Specifically, o
\Cref{sec:input-synthesis}
 describes the full steps of \brepair for this input.
 %Our \brepair algorithm
%%starts with the corrupted input and quickly
%finds the
%maximal parsable prefix using a binary search.
%For example, in \Cref{fig:bad-json-input},
Specifically, \brepair first finds the maximal parsable
prefix to be \letterboxed{\{\ "name":\ "Dave"\ }, and
determines that the \textit{parse boundary}, i.e., the
shift from \emph{incomplete} to \emph{incorrect}
happens at boundary %as %\\
\letterboxed{\{\ "name":\ "Dave"\ "}. % where the parser response shifts from
%\emph{incomplete} to \emph{incorrect}.
%At this point,
Next, \brepair applies \textit{deletion}, or \textit{insertion} of characters
in order.
In this case, the \letterboxed{"} is deleted first, but the
resulting input %\\
\letterboxed{\{\ "name":\ "Dave"\ age":\ 42 \}} %is
%checked to
%see if it resulted in increasing the size of the
does not increase the
maximum parsable prefix. %Here, we see that the maximal parse boundary remains the same.
Hence, \brepair next
attempts to insert a character,
%. There are 128 printable characters in ASCII,
%out of which,
eventually determining that
only space characters and
comma (\letterboxed{,}) can be inserted here,
resulting in an increase of the maximal parse prefix. Out of these two insertion candidates, inserting comma results in the maximum advancement of the parse boundary, resulting in the
valid repair: %newly constructed string: %\\
\letterboxed{\{\ "name":\ "Dave"\ ,"age":\ 42\ \}}. %,
%which is accepted as a valid repair.
\done{In Table~1, the comma is inserted after the quote; here, it is inserted after the space -- AZ}

 %Further, the list of repairs it suggests will include the
%original string before corruption.
%\\
%\\
%\noindent\textbf{Contributions.}
%\todo{we need to emphasize the main novelty here, i.e, first to do X, to the best of our knowledge }
In this work, we % paper %We
show that invalid inputs with rich structure can be repaired adequately if
the input processor can provide at least some indication of progress.
That is, as with \ddmax, we require the input processor to indicate if the input
is valid.  However, when given an invalid input, we \textit{require} the input processor
to indicate whether the input is merely \emph{incomplete} (that is, the input
is a prefix of a valid input) or is \emph{incorrect} (that is, no suffix to
this input will result in a valid input).\footnote{This requirement is
%not hard to accomplish. In most cases,
provided by most parsers. %,
%already
%provide precise information as to the point at which the parse failed,
%which
%can be used directly in our algorithm.
In the cases where parsers do not provide
this information, it can be obtained
%obtaining it %this information 
by external instrumentation 
%ispossible 
as demonstrated by Bj\"orn et al.~\cite{mathis2019parser}, or 
% If
%the program is hard to instrument,
by modifying the parser to provide such failure feedback, as demonstrated in this work. 
% to the parser. % that we require
%is trivial.
%not difficult.
} \revise{
We note that satisfying the requirements for \approach does not require using a
formal grammar for parsing, and indeed, there are several systems that
implement handwritten parsers that satisfy \approach
constraints~\cite{eaton2021parser}.}
% Parser generators vs. handwritten parsers: surveying major language implementations in 2021
% https://notes.eatonphil.com/parser-generators-vs-handwritten-parsers-survey-2021.html
%, i.e.,
%determining the incompleteness or incorrectness of an input
%this information can be provided by a parser without a grammar. 
%Even though a parser  often times employ a formal grammar for validity checks, 
%to determine the incompleteness or incorrectness of an input, %However, 
%There are several systems satisfying our constraints by implementing handwritten parsers without formal grammar. For example, many implementations of JSON are handwritten without a formal grammar~\todo{cite}.
%In our evaluation, u

%\recheck{
To the best of our knowledge, \approach is the \textit{first} approach to effectively repair invalid rich inputs without an input specification 
%(e.g., grammar) 
or program analysis.
%: It repairs invalid inputs %achieves this 
%by synergistically combining \textit{lightweight failure feedback and input synthesis}. 
%We evaluate \brepair using 806 (real-world) invalid inputs belonging to four well-known input formats (e.g., cJSON and TinyC). 
%Our evaluation results show that %given input processors that can accurately
%%signal \emph{incompleteness} or \emph{incorrectness} and % when given
%%invalid inputs,
%\textit{\brepair has a high (91\%) data recovery rate}. 
%%It recovers 91\% of input data 
%%and it repairs 77\% of invalid inputs and, within four minutes.}
%% and recovers 91% of input data,
%% (77\% repairs within four minutes), and 
%It is also up to 35\% more effective than \ddmax, the state-of-the-art input repair method.
%%and \recheck{X\% more efficient} %can provide repairs that are more maximal
%%than \ddmax.
%%, and correctly
%%pass the parser.
%%Overall, t
This paper makes the following %technical
contributions: %\todo{we need to stress some results here}:
\begin{itemize}
%\todo{discuss the repair to subset vs. insertion, refer to \Cref{sec:input-synthesis}}.
\item \textbf{Repair via Input Synthesis.} \brepair expands the repertoire for input repair to \emph{deletion},
\emph{insertion}, and \emph{modification}.
Unlike \ddmax, which is limited to \emph{deletion} of input
fragments, \approach can repair errors 
% which makes it unsuitable for repairing errors 
due to omission. % and corruption. 
\item \textbf{Lightweight Failure Feedback.} \brepair is the \textit{first} technique
to demonstrate %hat shows 
how to use lightweight parser-error-feedback (e.g., \textit{incomplete} checks) for input repair.
%We propose \brepair which does not have these limitations. \todo{refer to \Cref{sec:brepair}}
 %	\item \textbf{}	
 %  \item We identify one of the major limitations to \ddmax{}---the limitation due to rich struture, which can be fixed only by requiring additional constraints.
 %  \item We provide two bug fixes to \ddmax.

\item \textbf{Repairing Rich Input Structures.} We show that our
\brepair technique can repair inputs with rich structure, including multiple faults and large spans. This is a major limitation of the state-of-the-art \ddmax. 
%is its inability to handle \textit{inputs with rich structure}. 

\item \textbf{Empirical Evaluation.} We evaluate \brepair using 806 (real-world) invalid inputs belonging to four well-known input formats (e.g., cJSON and TinyC). 
Our evaluation results show that %given input processors that can accurately
%signal \emph{incompleteness} or \emph{incorrectness} and % when given
%invalid inputs,
\textit{\brepair has a high (91\%) data recovery rate}. 
%It recovers 91\% of input data 
%and it repairs 77\% of invalid inputs and, within four minutes.}
% and recovers 91% of input data,
% (77\% repairs within four minutes), and 
It is up to 35\% more effective than \ddmax.
%, the state-of-the-art input repair method.
%and \recheck{X\% more efficient} %can provide repairs that are more maximal
%than \ddmax.
%, and correctly
%pass the parser.
%Overall, t
%\item \textbf{}

\end{itemize}

The remainder of this paper is structured as follows: \Cref{sec:rich-input-structures} highlights the limitations of the state-of-the-art input repair method (\ddmax) and illustrates how \brepair overcomes these limitations. In \Cref{sec:input-synthesis} we describe how \brepair conducts input synthesis, and \Cref{sec:brepair}  describes the \brepair algorithm. We describe our experimental setup and findings in \Cref{sec:experimental-setup} and \Cref{sec:results}. We address the limitations %and threats to the validity 
of this work in \Cref{sec:threats}, and discuss related work in \Cref{sec:related_work}. Finally, we conclude this paper 
% with the discussion of related work and conclusions in \Cref{sec:related_work} and 
with the discussion of future work in \Cref{sec:conclusion}. 
%, respectively.

%\todo{describe sections/order}

\begin{figure*}[t]
\begin{boxedminipage}{\textwidth}
\smallskip
\ \begin{minipage}{0.9\textwidth}
\subsection*{Maximizing Delta Debugging Algorithm}
\medskip

Let $\test$ and $\cfail$ be given such that $\test(\emptyset) = \pass \land
\test(\cfail) = \fail$ hold.

The goal is to find $\dpass = \ddmax(\cfail)$ such that $\dpass \subset \cfail$, $\test(\dpass) = \pass$, and~$\Delta = \cfail - \dpass$ is 1-minimal.

The \emph{maximizing Delta Debugging algorithm} $\ddmax(c)$ is
\begin{align*}
\ddmax(\cfail) &= \ddmax_2(\emptyset, 2) \quad \text{where} \\
\ddmax_2(\dpass, n) &=
\begin{cases}
    %if len(minus(CX_I, cprime_y)) == 1: return cprime_y
  \textcolor{red}{\dpass} & \text{\textcolor{red}{\hphantom{else }if $|\cfail - \dpass| = 1$} (``base case$^{\textcolor{red}{a}}$'')} \\
\ddmax_2(\cfail - \Delta_i, 2) & \text{else if $\exists i \in \{1, \dots, n\} \cdot \test(\cfail - \Delta_i) = \pass$ (``increase to complement'')} \\
\ddmax_2\bigl(\dpass \cup \Delta_i, \max(n - 1, 2)\bigr) &
\text{else if $\exists i \in \{1, \dots, n\} \cdot \test(\dpass \cup \Delta_i) = \pass$ (``increase to subset'')} \\
  \ddmax_2\bigl(\dpass, \min(|\cfail \textcolor{red}{- \dpass}|, 2n)\bigr) & \text{else if $n < |\cfail - \dpass|$ (``increase granularity$^{\text{\textcolor{red}{b}}}$'')} \\
\dpass & \text{otherwise (``done'').}
\end{cases}
\end{align*}
where $\Delta = \cfail - \dpass = \Delta_1 \cup \Delta_2 \cup \dots \cup \Delta_n$, all
$\Delta_i$ are pairwise disjoint, and $\forall \Delta_i \cdot |\Delta_i| \approx |\cfail - \dpass| / n$
holds.

The recursion invariant (and thus precondition) for $\ddmax_2$ is
$\test(\dpass) = \pass \land n \leq |\Delta|$.\\
\textcolor{red}{a}: Bugfix: This base case is necessary to ensure that repairing JSON input \letterboxed{1*1} does not violate the invariant $n \leq |\Delta|$.\\
\textcolor{red}{b}: Bugfix: We should look for minimum of the remaining so that invariant $n \leq |\Delta|$ is not violated for JSON input \letterboxed{\{*"":2\}}.
\end{minipage}
\end{boxedminipage}
%\vspace{-0.5\baselineskip}
\caption{Modified Maximizing Lexical Delta Debugging algorithm, extended from Kirschner et al.~\cite{kirschner2020debugging}}
%\vspace{-0.5\baselineskip}
\label{fig:ddmax}
\end{figure*}

\section{Rich Input Structures} %Limitations of \ddmax}
\label{sec:rich-input-structures}
%\todo{discuss first (two) examples}
%---
% Example for Span Problem
%---
% \begin{figure}
% \begin{center}
% \letterboxed{\{\ "item":\ "Apple",\ "price":\ ***3.45\ \}}
% \end{center}
% %\vspace{-0.6\baselineskip}
%   \caption{\ddmax repairs this JSON to \mbox{``\<\ \ \ \ 3.45>''}}
%    %\Cref{fig:bad-json-input-asterisks}. This is very
% %similar to Kirchner et al.~\cite[Figure 1]{kirschner2020debugging}
% \label{fig:bad-json-input-asterisks}
%
% \end{figure}
Let us illustrate the limitations of the state-of-the-art input repair method (\ddmax) and how our approach (\brepair) addresses these limitations.
\Cref{fig:ddmax} provides a modified version of the lexical \ddmax algorithm presented in Kirschner et al.~\cite{kirschner2020debugging}. A major limitation of lexical \ddmax is that it results in sub-optimal repairs when the invalid input contains \textit{rich structures}, e.g., \textit{multiple faults}.
%needs to conform to specifications such as context-free grammars.
In the following, we discuss these limitations.
% specific instances when \ddmax can repair an input suboptimally.

\subsection{Limitations due to multiple faults}
A pattern of failure of \ddmax occurs when \ddmax is given an input with
multiple errors.  For example, consider the JSON input \letterboxed{[*]+}.
Here, the JSON string is invalid because of two invalid characters that are
non-contiguous. The operation of \ddmax (\Cref{fig:ddmax}) proceeds as follows:

\begin{enumerate}
\item The operation starts with $\ddmax_2(\emptyset, 2)$
\item  $|\cfail - \emptyset| \ne 1$. Hence, the base case does not apply
\item Increase to complement:\\
$\cfail - \Delta_1$= \letterboxed{]+} \fail \\
$\cfail - \Delta_2$= \letterboxed{[*} \fail \\
\item Increase to subset:\\
$\emptyset \cup \Delta_1$=\letterboxed{[*} \fail \\
$\emptyset \cup \Delta_2$= \letterboxed{]+} \fail \\
\item Increase granularity: $n < |\cfail-\dpass|$ which is $2 < |\cfail-\emptyset|$ \pass \\
  Hence the next iteration is:
  $\ddmax_2\bigl(\emptyset, 4)\bigr) $
\item  $|\cfail - \emptyset| \ne 1$. Hence, the base case does not apply.
\item Increase to complement:\\
$\cfail - \Delta_1$= \letterboxed{*]+} \fail \\
$\cfail - \Delta_2$= \letterboxed{[]+} \fail \\
$\cfail - \Delta_3$= \letterboxed{[*+} \fail \\
$\cfail - \Delta_4$= \letterboxed{[*]} \fail \\
\item Increase to subset:\\
$\emptyset \cup \Delta_1$=\letterboxed{[} \fail \\
$\emptyset \cup \Delta_2$=\letterboxed{*} \fail \\
$\emptyset \cup \Delta_3$=\letterboxed{]} \fail \\
$\emptyset \cup \Delta_4$=\letterboxed{+} \fail \\
\item Increase granularity: $4 < 4$ \fail\\
\item The solution is $\emptyset$.
\end{enumerate}

That is, \ddmax is unable to optimally repair inputs of this kind which
contains multiple errors. While in this example, the data loss that occurred
may seem somewhat limited, this need not always be the case. A similar example
is given in \Cref{tab:ddmaxlimitations}.
Here, \letterboxed{\{"ABCD":[*"1,2,3,4,5,6"]*\}} contains two distinct
corruptions.
As in the previous case, \ddmax attempts to fix this input by dividing it into
smaller and smaller fragments, none of which isolates an error that when
removed, results in the solution \letterboxed{123456} with %which clearly has suffered
significant data loss, including the loss of structure and change in
input fragment type from string to number.
Hence, \ddmax cannot effectively repair inputs containing multiple faults.

\subsection{Effect of input decomposition}
Unfortunately \ddmax can produce non-optimal results even when the errors are
contiguous, and hence considered \emph{single} by \ddmax. The problem happens
when the corruption in the input interacts with the fragment decomposition
algorithm of \ddmax.
As an example, consider a variant of the previous input: \letterboxed{[*+]}.
The JSON string is invalid here because it contains two invalid characters
which are contiguous.
The operation of \ddmax (\Cref{fig:ddmax}) is as follows:

\begin{enumerate}
\item The operation starts with $\ddmax_2(\emptyset, 2)$
\item  $|\cfail - \emptyset| \ne 1$. Hence, the base case does not apply
\item Increase to complement:\\
$\cfail - \Delta_1$= \letterboxed{+]} \fail \\
$\cfail - \Delta_2$= \letterboxed{[*} \fail \\
\item Increase to subset:\\
$\emptyset \cup \Delta_1$=\letterboxed{[*} \fail \\
$\emptyset \cup \Delta_2$= \letterboxed{+]} \fail \\
\item Increase granularity: $n < |\cfail-\dpass|$ which is $2 < |\cfail-\emptyset|$ \pass \\
  Hence the next iteration is:
  $\ddmax_2\bigl(\emptyset, 4)\bigr) $
\item  $|\cfail - \emptyset| \ne 1$. Hence, the base case does not apply.
\item Increase to complement:\\
$\cfail - \Delta_1$= \letterboxed{*+]} \fail \\
$\cfail - \Delta_2$= \letterboxed{[+]} \fail \\
$\cfail - \Delta_3$= \letterboxed{[*]} \fail \\
$\cfail - \Delta_4$= \letterboxed{[*+} \fail \\
\item Increase to subset:\\
$\emptyset \cup \Delta_1$=\letterboxed{[} \fail \\
$\emptyset \cup \Delta_2$=\letterboxed{*} \fail \\
$\emptyset \cup \Delta_3$=\letterboxed{+} \fail \\
$\emptyset \cup \Delta_4$=\letterboxed{]} \fail \\
\item Increase granularity: $4 < 4$ \fail\\
\item The solution is $\emptyset$.
\end{enumerate}

That is, this particular invalid JSON string also cannot be repaired by \ddmax.
As in the previous case, the data loss can be severe.
Consider
\letterboxed{\{\ "item":\ "Apple",\ "price":\ ***3.45\ \}} in
\Cref{tab:ddmaxlimitations} which is similar to
Kirchner et al.~\cite[Figure 1]{kirschner2020debugging} but with an extra
\letterboxed{*}. \ddmax repairs this input to \letterboxed{\ \ \ \ 3.45},
resulting in data loss. % of information.

The problem here is that the successive partitions attempted by \ddmax fails to
isolate the failure causing fragment even though the fragment is contiguous.
That is, no single independent fragment is found, the removal of which results
in removal of the error. Hence, \ddmax keeps searching for smaller and smaller
fragments discarding larger and larger chunks of data.

\subsection{Discussion}
Why does \ddmax fail to repair these inputs? A major limitation of \ddmax is
that it is modeled on \ddmin, which is an effective tool for minimization of
failure inducing inputs. Given a failure inducing input, the idea of \ddmin is
to successively partition the input into smaller and smaller chunks, remove one
chunk at a time and check whether the remaining chunks are sufficient to
reproduce the failure. As this implies, a key assumption of \ddmin is that
we can actually remove such chunks independently. That is, if a chunk does not
contribute to the observed failure, it can be removed without affecting the
failure observed. Secondly, if multiple chunks independently cause
the same failure, only one chunk will be chosen, and minimized further.

The definition of \ddmax is a mirror of \ddmin. \ddmax starts with an empty
input that is assumed to be passing. Then, it partitions the input into chunks,
and tries to concatenate any of these chunks to the passing input, producing a
larger passing input. If after dividing the input into \<n> chunks, none of the
chunks could produce a passing input, it tries again by dividing the
input into \<2n> chunks.

As in the case of \ddmin, the \textit{unstated assumption} here is that if a chunk was
not responsible for the observed failure, it can be extracted independently of
other chunks and added to the passing input fragment without changing the
semantics. This particular \textit{assumption need not hold when we are dealing with
inputs that have a rich structure}. That is, \letterboxed{"1,2,3,4,5"} is very
different from \letterboxed{12345} even though a significant portion of the raw
characters from first is preserved in the second. Further, once such a
semantically changed fragment forms the seed of the passing fragment, due to
the constraints in the input structure, the remaining fragments from the
original will likely not combine with the seed fragment, resulting in further
data loss.

Although \ddmax will have no problems maximizing any inputs \textit{if  the input processor conforms to this constraint}, we note that this can be a rather \textit{strong
constraint in practice}.

%One of the ambiguities here is that it is not immediately
%clear what a repair means. For example the empty set $\emptyset$ is assumed to
%be passing. Hence, is $\emptyset$ a valid repair?

% \begin{figure}
% \begin{center}
% \letterboxed{\{\"ABCD":[*"1,2,3,4,5,6"]*\}}
% \end{center}
% %\vspace{-0.6\baselineskip}
% \caption{\ddmax repairs this JSON to ``\<123456>''}
% \label{fig:bad-json-input2}
% \end{figure}
%
% \begin{figure}
% \begin{center}
% \letterboxed{\{\ "name":\ "Daveage"\ \}}
% \end{center}
% %\vspace{-0.6\baselineskip}
% \caption{Failing input repaired with DDMax}
% \label{fig:bad-json-input-ddmax}
% \end{figure}

\subsection{Updates to \ddmax definition}
%\todo{discuss the bug fixes, base case and additional constraints here }
While evaluating \ddmax, we noticed two cases where the formal definition of
\ddmax was underspecified. These are noted in \Cref{fig:ddmax}. Specifically,
(1)~\ddmax requires the base case when $|\cfail-\cpass| = 1$. If not, \ddmax can
go into unbounded recursion on inputs such as the JSON input:
\letterboxed{1*1}. (2)~when increasing granularity, the size of the remaining
input should be considered rather than the size of the entire text. Not doing
this would cause an invariant fail for inputs such as \letterboxed{\{*"":2\}}.

\subsection{Repair of rich inputs with \brepair}
One of the strengths of \brepair is that it can effectively handle inputs with
rich structure. Consider the input to the JSON processor: \letterboxed{\{"ABCD":[*"1,2,3,4,5,6"]*\}}.
(For ease of explanation, let us consider only deletion as the
operation used.) Here, the procedure is as follows:

\begin{enumerate}
\item \brepair starts by executing a binary search for the boundary where the
input prefix changes from \emph{incomplete} to \emph{incorrect}. This is
obtained at index 10, providing the incomplete substring
\letterboxed{\{"ABCD":[}. %*"1,2,3,4,5,6"]*\}}.
\item \brepair then appends the next character \letterboxed{*} to the input,
resulting in \letterboxed{\{"ABCD":[*} %"1,2,3,4,5,6"]*\}}.
and observes the result. In this case, the JSON processor returns
\emph{incorrect}.
\item Hence, the newly added character is discarded, and the character at the
next index is appended, resulting in
\letterboxed{\{"ABCD":["}. %1,2,3,4,5,6"]*\}}.
This results in JSON processor responding \emph{incomplete}.
\item \brepair now appends the character in the next index, resulting in
\letterboxed{\{"ABCD":[1"} which again results in \emph{incomplete} from JSON
processor.
\item Proceeding in this fashion the input reaches \\
\letterboxed{\{"ABCD":["1,2,3,4,5,6"]*} %\}}.
at which point, we again have the response \emph{incorrect} from the JSON
processor. Hence, we discard this character, and try the next character,
resulting in
\letterboxed{\{"ABCD":["1,2,3,4,5,6"]\}}.
\item The JSON processor responds with \emph{complete}.
\end{enumerate}
This completes the repair of the given input. This demonstrates that \brepair has no problem repairing rich inputs containing multiple errors.

\section{Input Synthesis}
\label{sec:input-synthesis}
%\begin{figure}
%\begin{center}
%\letterboxed{\{\ "name":\ "Dave"\ "age":\ 42\ \}}
%\end{center}
%%\vspace{-0.6\baselineskip}
%\caption{Failing JSON input}
%\label{fig:bad-json-input}
%\end{figure}
%\begin{figure}
%\begin{center}
%\letterboxed{\{\ "name":\ "Daveage"\ \}}
%\end{center}
%%\vspace{-0.6\baselineskip}
%\caption{Failing input repaired with DDMax}
%\label{fig:bad-json-input-ddmax1}
%\end{figure}

%\todo{discuss importance of insertions and limitations of repair to subsets}
The second major limitation of
\ddmax is that the only operation in its toolbox is \emph{deletion} of input
fragments. Consider\\
\letterboxed{\{\ "name":\ "Dave"\ "age":\ 42\ \}}
%\Cref{fig:bad-json-deletion}.
%\begin{figure}
%\begin{center}
%\letterboxed{\{\ "fruits:\ ["Apple",\ "Orange",\ "Banana"]\}}
%\end{center}
%%\vspace{-0.6\baselineskip}
%\caption{Invalid JSON with non-optimal repair by \ddmax}
%\label{fig:bad-json-deletion}
%\end{figure}
%---
% Example for Insertion
%---
Here, there is a missing quote in the key. \ddmax repair of this string will
result in
\letterboxed{\ \ \ \ 42}.
The problem is that, deletion of fragments alone can lead to significant
corruption of information. In this instance, the availability of \emph{insertion}
could have repaired the input string to \\
\letterboxed{\{\ "name":\ "Dave",\ "age":\ 42\ \}}. Unfortunately, because
\ddmax is unable to synthesize any framents, opportunities for repair can be
missed.

The \brepair algorithm on the other hand, follows in this fashion.
\begin{enumerate}
\item \brepair algorithm starts with the corrupted input and quickly finds the
maximal parsable prefix using a binary search: \\
\letterboxed{\{\ "name":\ "Dave"\ }.
\item At this point, \brepair applies \emph{deletion}, \emph{insertion}, or
\emph{modification} of characters in order.
In this case, the \letterboxed{"} is deleted first, resulting in: \\
\letterboxed{\{\ "name":\ "Dave"\ age":\ 42 \}}.
\item The JSON parser responds with \emph{incorrect} for this input.
\item \brepair next attempts to insert a character. Say we tried to insert
\letterboxed{1}. This results in: \\
\letterboxed{\{\ "name":\ "Dave"\ 1"age":\ 42 \}}.
\item The JSON parser responds with \emph{incorrect} for this input.
\item Indeed, only space characters and comma (\letterboxed{,}) can be
inserted here, resulting in \emph{incomplete} from the JSON parser.
\item Inserting the \letterboxed{,} results in a new input: \\
\letterboxed{\{\ "name":\ "Dave"\ ,"age":\ 42\ \}}
\item This is accepted as a valid repair.
\end{enumerate}

That is, the ability of \brepair to synthesize characters for repair can lead
to more effective repairs.

\section{\approach}
\label{sec:brepair}
The main strength of \brepair is its repertoire of repairs.
The input string can be modified by \emph{deleting} any character
(\Cref{lst:repairsdelete}), or
\emph{inserting} a character at any index (\Cref{lst:repairsdelete}).
In the listings below, \<Repair> is a tuple that allows
accessing its first argument with \<inputstr> and its second argument with
\<boundary>.

We define a few terms before we start:
\begin{description}
\item[valid substring.] The \emph{valid substring} of a string is the maximal prefix where the parser returns
\emph{incomplete}.
\item[boundary.] The \emph{boundary} or \emph{parse boundary} is the index of
the first character after the valid substring that results in
\emph{incorrect} response from the parser, or one past the end if the input is
\emph{incomplete}.
\item[repair.] A repair is a single modification (deletion or insertion of a
single character) made on an input string.
\item[repair thread.] A \emph{repair thread} is is a set of repairs made on an
input string. A repair thread has a single \emph{boundary} and a corresponding
\emph{valid prefix}.
\end{description}

The \emph{deletion} algorithm is simple. When a parse error is observed,
the character that caused the error is present at the \<boundary> value.
That is, we know that \<inputstr[:boundary]> parses correctly. Hence, for
deletion, we produce a string without the character at the \<boundary>.
\begin{lstlisting}[caption=\brepair repairs,label={lst:repairsdelete}]
def apply_delete(item):
    inp = item.inputstr[:item.boundary] +
          item.inputstr[item.boundary + 1:]
    return extend_deleted_item(Repair(inp, item.boundary))
\end{lstlisting}

For \emph{insertion}, the algorithm is more involved. The problem is that we need to
handle corruption in inputs such as \<"12345mystring"> which is used as input
to a JSON parser. Consider what happens when the first quote is deleted. In this
case, we will get the first parse error at \<m>. But plainly, the best repair
is before the first parse error. The problem is that insertions at random
points in the prefix is very costly, resulting in $|\alpha| \times |S|$
modifications where $|\alpha|$ is the number of alphabets (characters) the
language has, and $|S|$ is the length of the prefix.
Hence, we allow the user to choose to toggle this option with
\<LAST\_INSERT\_ONLY>. If the option is false, we will attempt repairs at any
point in the prefix. If it is true, we will only attempt insertions at the end
of the prefix.
\begin{lstlisting}[caption=\brepair repairs,label={lst:repairs}]
def insert_at(item, k, i, suffix)
    v = (item.inputstr[:k] + i +
        item.inputstr[k:item.boundary] + suffix)
    new_item = Repair(v, k, mask='%s_I%d' % (item.mask, k))
    ie = extend_inserted_item(new_item)
    if ie.boundary > k:
       return ie
    return None

def insert_char(item, i):
    suffix = item.inputstr[item.boundary:]
    return_lst = []
    if LAST_INSERT_ONLY:
        v = insert_at(item, item.boundary, i, suffix)
        if v is not None: return_lst.append(v)
    else:
        for k in range(item.boundary):
            v = insert_at(item, k, i, suffix)
            if v is not None: return_lst.append(v)
    return return_lst

def apply_insert(item):
    new_items = []
    for i in CHARACTERS:
        items = item.insert_char(i)
        if items:
            new_items.extend(items)
    return new_items
\end{lstlisting}

\Cref{lst:brepair} shows how \brepair is invoked. The input string is passed to
\<repair()>.
\begin{lstlisting}[caption=\brepair initial search,label={lst:brepair}]
def repair(inp):
    boundary = binary_search(inp)
    return find_fixes(inp, boundary)
\end{lstlisting}
This function does a \emph{binary search} (\Cref{lst:bsearch}) on the
argument string looking for the parse boundary where the return value changes
from \emph{incomplete} to \emph{incorrect}. That is, if \<binary\_search>
returns an index \<n>, then \<inputstr[:boundary]> is a valid prefix, and
\<inputstr[boundary]> is the error causing character if one exists or the
\<inputstr> is \emph{incomplete}.
\begin{lstlisting}[caption=binary search,label={lst:bsearch}]
def binary_search(inputstr, left=0, right=len(inputstr)-1):
    if not inputstr: return left
    if is_incomplete(Repair(inputstr, right)):
        return len(inputstr)-1
    while left + 1 < right:
        middle = (left + right) // 2
        if is_incomplete(Repair(inputstr, middle)):
            left = middle
        else:
            right = middle
    return left
\end{lstlisting}
The \<boundary> value is then passed
to \<find\_fixes()> (\Cref{lst:findfixes}).
\begin{lstlisting}[caption=find fixes,label={lst:findfixes}]
def find_fixes(inputval, boundary):
    next_items = [Repair(inputval, boundary)]
    while True:
        current_items, next_items = next_items, []
        completed = []
        for item in sample_items_by_mask(current_items)
            for i in repair_and_extend(item)
                next_items.append(i)
                if i.is_complete(): completed.append(i)
        if completed: return completed
\end{lstlisting}
The function \<find\_fixes()> takes in the input string and the parse boundary.
It then generates a set of repair threads, out of which a few are chosen for
continuation (\Cref{lst:sampling}).
Then, each repair thread is processed individually.
On each thread, a set of repairs (\Cref{lst:repairsdelete}, \Cref{lst:repairs})
are applied. If the repair succeeds,
the string can use more characters from the pending suffix resulting in a
larger valid prefix string, but with a larger edit distance. This procedure is
repeated until the prefix string is marked \emph{complete} by the input
processor.

The sampling procedure attempts to discard redundant repairs so that the number
of simultaneous threads we have to maintain does not grow unbounded. While doing
that, it ensures that no unique repairs are discarded. For
example, given a JSON fragment \<[1,2> a repair of \<[1,23>, \<[1,24>
and \<[1,25> are redunant but \<[1,2,> is unique because the last repair
represents a change in semantics for the parser. The key insight here is to
look at the \emph{extension} of a given string to classify whether it is unique
or not. That is, given a string \<[1,2"x"]>, after repair of \<[1,23>, the next
extension of the string is likely to be a comma or a digit insertion. However,
after \<[1,2,> the pending suffix can be used to extend the string resulting in
\<[1,2,"x"]>.
%\recheck{
    Hence, we use the kind of repairs conducted on a string, the
length of the prefix, as well as the last character added as the
uniqueness indicator.
%}
%\todo{@Lukas: does the new implementation contain this check? I cant remember this check in the old implementation of \approach?} - Yes, it contains those checks: https://projects.cispa.saarland/lukas.kirschner/bfuzzerrepairer/-/blob/main/project/src/main/java/bfuzzerrepairer/program/repairer/brepair/BRepairRepairer.java#L628

Unfortunately, using all repairs even after eliminating all redundant repairs
can be rather time consuming. If this is the case, one can limit the repair
threads to the best performing ones in terms of the parse boundary by using
\<filter\_best()>.

\begin{lstlisting}[caption=Sampling,label={lst:sampling}]
def sample_items_by_mask(items):
    masks = {}
    for i in items:
        key = (i.mask, i.boundary, i.inputstr[i.boundary-1])
        if i.mask not in masks: masks[key] = []
        masks[key].append(i)

    sampled = []
    for key in masks:
        if len(masks[key]) < MAX_NUM_PER_MASK:
            res = masks[key]
        else:
            res = random.sample(masks[key], MAX_NUM_PER_MASK)
        sampled.extend(res)
    return filter_best(sampled)

def filter_best(items):
    if MAX_SIMULTANIOUS_CORRECTIONS < 0: return items
    boundaries = unique_sorted_boundaries(items)
    return [i for i in items if i.boundary in
            boundaries[:MAX_SIMULTANIOUS_CORRECTIONS]]
\end{lstlisting}

\begin{figure}
\newlength\nodedst\setlength\nodedst{.5cm}
\newlength\ndist\setlength\ndist{.1\nodedst}
\def\arrowsep{0.20cm}
\begin{tikzpicture}[node distance=\nodedst]
    \node[blockdiagramblock] (search) {Find parse boundary (Search)};
    \node[blockdiagramblock,below=of search] (repair) {Apply Repairs};
    \draw[blockdiagramarrow] (search) -- (repair); %node[simpletext,pos=.5,right,font=\scriptsize] {}; % PushPQ
    \node[blockdiagramblock,below=of repair] (extend) {\mbox{Extend Threads}};
    \draw[blockdiagramarrow] (repair) -- (extend);
    \node[blockdiagramblock,right=of extend] (isvalid) {\mbox{Valid Input?}};
    \node[simpletext,below=.3 of isvalid] (return) {\mbox{Return input}};
    \node[blockdiagramblock,above=of isvalid] (select) {\mbox{Select Threads}};
    \draw[blockdiagramarrow] (extend) -- (isvalid);
    \draw[blockdiagramarrow] (isvalid) -- (return) node[simpletext,pos=.6,right] {\pass};
    \draw[blockdiagramarrow] (isvalid) -- (select);
    %\node[simpletext,right=.5 of isvalid.east,inner sep=1pt] (pqins1) {increase bound,\\push to PQ};
    %\draw[blockdiagramarrow,dashed] (isvalid) -- (pqins1) node[pos=.5,above] {\pass};
    %\node[simpletext,right=.5 of select.east,inner sep=1pt] (pqins2) {increase bound,\\push to PQ};
    %\draw[blockdiagramarrow,dashed] (select) -- (pqins2) node[pos=.5,above] {\pass};
    \draw[blockdiagramarrow] (select) to[out=180,in=0] (repair);
    \pic[] at (search.south east) {numbering=1};
    \pic[] at ([yshift=-3mm]isvalid.south east) {numbering=5};
    \pic[] at (extend.south east) {numbering=3};
    \pic[] at (repair.south east) {numbering=2};
    \pic[] at (select.south east) {numbering=4};
    %TODO
\end{tikzpicture}
% \vspace*{-0.1in}
    \caption{Work flow of \brepair}
    \label{fig:brepair_flowchart}
    % \vspace*{-0.1in}
\end{figure}
\newcommand{\refnumber}[1]{\hyperref[fig:brepair_flowchart]{Step~#1}}

After each repair, the new string is checked  to see if
the string results in \emph{incomplete} rather than \emph{incorrect}. If the
string results in \emph{incomplete}, a new parse boundary is found by
repeatedly extending the string with a pending character from the remaining
suffix in the input string. The small complication here is that different
search algorithms are more suitable for deletion and insertion. With deletion,
we have removed the error causing character, so the next parse error may
be far away. Hence, we use binary search to find the next parse error (\Cref{lst:extenditem}).
\begin{lstlisting}[caption=Extend the boundary,label={lst:extenditem}]
def extend_deleted_item(item):
    return bsearch_extend_item(item)

def bsearch_extend_item(item):
    bs = binary_search(item.inputstr, left=item.boundary)
    if bs >= len(item.inputstr):
        item.boundary = bs
        return item
    item.boundary = bs
    return item
\end{lstlisting}
In the case of insert, however, in most cases, the character being inserted,
and the character that originally caused a parse error can still cause a parse
error. Hence, we apply linear search instead (\Cref{lst:extenditeminsert}).
\begin{lstlisting}[caption=Extend the boundary,label={lst:extenditeminsert}]
def extend_inserted_item(item):
    return lsearch_extend_item(item, nxt=1)

def lsearch_extend_item(item, nxt=1):
    while True:
        if (item.boundary + nxt) > len(item.inputstr):
            item.boundary = item.boundary + nxt - 1
            return item
        s = Repair(item.inputstr, item.boundary + nxt)
        if is_incomplete(s):
            nxt += 1
            continue
        if is_incorrect(s):
            item.boundary = item.boundary + nxt - 1
            return item
\end{lstlisting}
If this succeeds, the parse boundary after the repair
would be larger than the old parse boundary. Hence, if the parse boundary has
increased (\Cref{lst:repairandextend}), then
%. If the boundary has increased,
the
repair is saved. If not, the repair is discarded.
\begin{lstlisting}[caption=Repair and extend,label={lst:repairandextend}]
def repair_and_extend(item):
    return [apply_delete(item)] + apply_insert(item)
\end{lstlisting}
The output from \<find\_fixes()> (\Cref{lst:findfixes}) is a set of repair
threads with the least number of repairs from the passed input string.

%The \brepair algorithm works as follows.
%We start with the assumption that 
\approach assumes the parser correctly signals
\emph{incomplete} for incomplete inputs, \emph{incorrect} for other invalid
inputs, and \emph{complete} if the input was valid. 
Given this, \brepair algorithm works as follows (\Cref{fig:brepair_flowchart}):
%\todo{we need to sync these steps with the workflow diagram (\autoref{fig:brepair_flowchart}), and also name/number steps 1,2, ...}

\begin{enumerate}
\item \emph{Binary search.} Given any corrupt input, \brepair starts by a
binary search of the input to determine the parse boundary. This then is used
to construct the first repair thread, with boundary set to the binary search
result, and repairs set to empty.

\item \emph{Repair.} Starting with any existing repair thread, \brepair applies
deletion and insertion repairs. A single deletion results in a single repair
thread which is an extenion of the original repair thread.
An insertion however, results in multiple repair threads corresponding to the
number of characters.
\item \emph{Extention.} For each thread that results from repair, extend the thread
until the new parse boundary. For threads resulting from insertion, keep only
those threads that results in a diffrent parse boundary from the old.
\item \emph{Selection.} We remove any redundant repair threads, and choose the best threads if a
selection criterion is supplied.

\item \emph{Final Fix.} We then iteratively choose each thread and continue making repairs
until the parser output at the boundary changes from \emph{incomplete} to
\emph{complete}.
\end{enumerate}

The output of the algorithm is a set of repair threads each of which fixes the
input using a set of repairs. The repair threads are sorted in the order of
least number of repairs required to fix the input.

%\noindent
%\textbf{Complexity:}\todo{what is the space/time complexity of the \approach, and how does this compare to that of \ddmax. }

%\section{Approach}

%
%\hline
%Example & \ddmax repair & \approach repair & \ddmax limitation \\
%\hline
%\letterboxed{\{\ "item":\ "Apple",\ "price":\ ***3.45\}} & \letterboxed{\ \ \ \
%3.45} & \letterboxed{\{\ "item":\ "Apple",\ "price":\ 3.45\}} & Rich structure
%(spans) \\
%\letterboxed{\{\"ABCD":[*"1,2,3,4,5,6"]*\}} &
%\letterboxed{123456} &
%\letterboxed{\{\"ABCD":["1,2,3,4,5,6"]\}} &
%Rich Structure (multiple-faults) \\
%\letterboxed{\{\ "name":\ "Dave"\ "age":\ 42\ \}} &
%\letterboxed{\{\ "name":\ "Daveage":\ 42\ \}} &
%\letterboxed{\{\ "name":\ "Dave",\ "age":\ 42\ \}} &

%
%\begin{table*}[!tbp]\centering
%\caption{Debugging Richly Structured Inputs: \approach versus \ddmax}
%\begin{tabular}{|l | c | r | r | r | r |}
%\hline
%\textbf{Challenges} & \textbf{Sample Input}  &  \textbf{\ddmax Repair} & \textbf{\approach Repair} & \textbf{\approach Strength}  \\
%\hline
%\textbf{Search Span} %(Rich Structure)}
%&  \{\ "item":\ "Apple",\ "price":\ ***3.45\} & & & \\
%\textbf{Multiple faults} & \{\ "ABCD":[*"1,2,3,4,5,6"]*\} & & & \\
%\textbf{Repair to subsets} & \{\ "name":\ "Dave",\ "age":\ 42\ \} & & & \\
%\textbf{Rich Structure} & & & & \\
%\textbf{Limited Repair} & & & & \\
%\textbf{Failure Feedback} & & & & \\
%\hline
%\end{tabular}
%\label{tab:challenges-inp-debug}
%%\vspace{-0.5cm}
%\end{table*}

\begin{table}[!tbp]\centering
\caption{Subject programs used in the evaluation}
\begin{tabular}{|l | r | l | l | l | l |}
\hline
\textbf{Name} & \textbf{LOC} & \textbf{Lang.} & \textbf{1{st} Commit} & \textbf{Last Commit} \\
\hline
\textbf{INI} & 382 & C & Jul 2009 & Jan 2022\\
\textbf{cJSON} & 3062 & C & Aug 2009 & Jan 2022\\
\textbf{SExpParser} & 656 & C & Sep 2016 & Sep 2016\\
\textbf{TinyC} & 375 & C & 2001 & Apr 2018\\
\hline
\end{tabular}
\label{tab:subject-programs}
%\vspace{-0.5cm}
\end{table}

\section{Experimental Setup}
\label{sec:experimental-setup}

%TODO This is copied from ddmax, change?

This section describes the experimental setup of this work.

\begin{table}[!tbp]\centering
\caption{Details of Invalid 
%Failure-inducing 
Inputs}
\begin{tabular}{|l | r | r | r | r |}
\hline
&  \multicolumn{4}{c|}{\textbf{Number of Invalid Inputs}}  \\
\textbf{Type of Invalid Inputs} & \textbf{INI} & \textbf{cJSON} & \textbf{SExp} & \textbf{TinyC} \\
\hline
\textbf{Real-World Inputs} & 101 & 107 & 50 & 148 \\
\textbf{Single Mutation} & 50 & 50 & 50 & 50 \\
\textbf{Multiple Mutations} & 50 & 50 & 50 & 50 \\
\hline
\textbf{Total } (806) & 201 & 207 & 150 & 248 \\
\hline
\end{tabular}
\label{tab:input-details}
%\vspace{-0.5cm}
\end{table}

\begin{table*}[!tbp]\centering
\caption{Data Recovery (file size difference) and Data Loss  (Levenshtein distance) of \approach vs. %the best-performing state-of-the-art methods, i.e. 
lexical (Lex.) \ddmax and syntactic (Syn.) \ddmax. 
The highest data recovery and lowest data loss are in bold, percentage improvement (Impr.) %or reduction (red.) 
of \approach over the best baseline which are significant (i.e., greater than five percent ($>$5\%)) are also in bold.
}
%\vspace{-0.5cm}
\begin{tabular}{|l | c |  r  r  r  r  | c |  r  r  r  r |}
\hline
 &  \multicolumn{5}{c|}{\textbf{\% Data Recovered by Approach}} &  \multicolumn{5}{c|}{\textbf{Average Data Loss }}  \\
&  \multicolumn{1}{c|}{\textbf{ALL}} & \multicolumn{4}{c|}{\textbf{Input Format}}  &  \multicolumn{1}{c|}{\textbf{ALL}} & \multicolumn{4}{c|}{\textbf{Input Format}}  \\
\textbf{Techniques} & \textbf{Average} & \textbf{INI} & \textbf{cJSON} & \textbf{SExp} & \textbf{TinyC} & \textbf{Average} & \textbf{INI} & \textbf{cJSON} & \textbf{SExp} & \textbf{TinyC} \\
\hline
\textbf{Lex. ddmax} & 81\%  & 85.8\% & 82.2\%	 & 99.2\%	& 47.7\% & 18.0 & {10.2} &	61.6 &	\textbf{6.2} & 7.5 \\			
\textbf{Syn. ddmax} & 74\%  &  70.6\% & 81.5\%  & 97.0\%	& 47.0\% & 119.6 &  258.1 & 82.7 &	76.6 &	33.3 \\	
\hline
\textbf{\approach} & \textbf{91\%}  &  \textbf{85.9\%} & \textbf{98.5\%} & \textbf{100.6\%} & \textbf{82.1\%} & \textbf{10.6}  & \textbf{9.5} &	\textbf{28.7} &	7.4 & \textbf{1.6} \\
\hline
\textbf{Impr. vs. Lex.} & 12\% &  0.1\%	& \textbf{19.8\%}	& 1.4\%	& \textbf{72.1\%} & \textbf{70\%} &  \textbf{8\%} & \textbf{115\%} & -16\% & \textbf{357\%} \\
\textbf{Impr. vs. Syn.} & \textbf{23\%}  &  \textbf{21.6\%} 	& \textbf{20.8\%}	& 3.7\%	& \textbf{74.7\% }  & \textbf{1030\%} & \textbf{2629\%} &	\textbf{188\%} & \textbf{936\%} &\textbf{ 1930\%} \\
\hline
%\textbf{Total \#Repairs} & (\%) & &  & 	& \\		
%\hline
\end{tabular}
\label{tab:data-recovery-and-loss}
%\vspace{-0.5cm}
\end{table*}

\noindent\textbf{Research Questions:} 
We investigate the \textit{data recovery} (RQ1), \textit{effectiveness} (RQ2), 
%\textit{diagnostic quality} (RQ3) 
and \textit{efficiency} (RQ3) of our approach using several well-designed experiments. %For each experiment, w
We also examine how \approach compares to four state-of-the-art techniques, 
namely the built-in repair of the programs (\textit{baseline}), error-recovery of \textit{ANTLR}, as well as 
%'s error recovery strategy, 
%lexical \ddmax 
lexical \ddmax and syntactic 
\ddmax. Specifically, we pose the following research questions:
\begin{description}[wide, labelwidth=!, labelindent=0pt]

\item[RQ1: Data Recovery and Data Loss.]
How much input data is recovered by \approach, and how much data is lost? %s is incurred?  
Does \approach recover as much data as 
the state-of-the-art methods?

  \item[RQ2: Effectiveness.]  %\& Efficiency:}
How effective is \approach in fixing invalid inputs? Is it as effective as the state-of-the-art methods?

\item[RQ3: Efficiency.] What is the efficiency (runtime) of
\approach? Is it as efficient as the state-of-the-art techniques?
\end{description}

\noindent\textbf{Subject Programs:} %As subjects w
We used four input formats and their corresponding programs. These are INI (INI), JSON (cJSON), S-Expressions (SExpParser) and TinyC (TinyC). Each program is moderately large (between 375 LOC to 3062 LOC), relatively mature (6 to 21 years old), and written in \<C>. Further details are provided in \Cref{tab:subject-programs}.

%\subsubsection*{\bf Test inputs}\todo{refactor this}
\noindent
\textbf{Test inputs:} 
\Cref{tab:input-details} provides details of the number of
real-world invalid inputs and mutated invalid inputs employed in our experiment.
%,for each input format. 
We evaluate our approach using 806 invalid input files. %.
%belonging to four different input formats, namely INI, cJSON, SExp, and TinyC. 
As test inputs, we crawled a large corpus of valid and invalid real-world files
from GitHub using the GitHub crawling API~\cite{githubapi}. 
%TODO add table with number of files?
In addition to real invalid inputs, for each format, we introduced a set of 100 artificially mutated (invalid) files
%that we mutated 
from 50 randomly selected valid real-world files.
Half of those mutated files contain a single mutated byte and the other half
contains multiple (two to 16) mutated bytes.
Each mutation can either be a byte-flip, an insertion or a deletion to 
%, which we believe to 
resemble real-world corruptions as close as possible
(e.g., bit rot on hard disks or transmission errors in network protocols).

\noindent\textbf{Metrics and Measures:} We employ the following metrics and measures to evaluate repair quality: 

\noindent\textit{a.) Number of Repaired Inputs:} We count the number of files repaired before
a four minute timeout for each repair method.
%\begin{description}[wide, labelwidth=!, labelindent=0pt]
 %of the tested repair programs.
 
\noindent\textit{b.) File Size Difference:} To determine the amount of data recovered by each approach, 
we evaluate the difference in \emph{file size} of
the \emph{recovered inputs} and the original
\emph{valid input}. %. This measurement tracks the effectiveness of 

\noindent\textit{c.) Edit Distance:} This is measured as the number of characters that
differs between the \emph{corrupt input} and the \emph{repaired input}.
%  \cite{levDistance}.

\noindent\textit{d.) Runtime} is the time taken for input repair for each method.

\noindent\textit{e.) Number of Program Runs:} To evaluate efficiency, we track the number of times the parser is executed by each approach. 
%during test experiments of input fragments. 
%we need to run the
%parser can provide a useful metric for evaluation. %Hence, we track this also.
%\end{description}

\noindent\textbf{State-of-the-art:}
In this work, we compare the performance of \approach to the following
techniques:
%\begin{description}[wide]
\noindent\textbf{(1) Baseline:} The built-in error-recovery technique of the subject programs.
 % In this work,  \recheck{we compare the effectiveness of the subject program to that of \approach (\textit{see} RQ1)}. Specifically, we measure the effectiveness of the built-in error recovery feature of the program by measuring the number of invalid input files that could be processed as valid inputs by a subject program without inducing a program failure, crash or timeout. In our evaluation, we compare the number of such repairs to that of \approach.
\noindent\textbf{(2) ANTLR:} This is the inbuilt error recovery strategy of the ANTLR parser generator when equipped with an input grammar specifying the allowed input structure~\cite[Automatic Error Recovery Strategy]{parr2013definitive}.
  %In this work, we compare the number of invalid inputs that could be processed by \textbf{ANTLR} to the number of repairs performed by \approach (\textit{see} RQ1).
\noindent\textbf{(3) \ddmax:} Kirschner et al.~\cite{kirschner2020debugging} proposed %This refers to the
two variants of the maximizing variant of the delta debugging algorithm (called \ddmax), %proposed by ,
namely \textit{lexical \ddmax} and \textit{syntactic \ddmax}.
    %For all research questions in this work, we compare the performance of \approach to both \textit{lexical} and \textit{syntactic} \ddmax, since \ddmax is the current best performing state-of-the-art input repair method.
%\item[\ddmin:] Zeller and Hildebrandt proposed
%a delta debugging (DD) algorithm designed to diagnose the root cause of a program failure %, \ddmin isolates a failure cause
%by minimizing a failure-inducing input to the minimal input subset that reproduces
%the observed program failure. In our evaluation of the diagnostic quality of \approach (\textit{see} RQ3), we compare the performance of \approach to \ddmin.
%\end{description}

%(a)  (\textit{baseline}), (b) the inbuilt error recovery strategy of the ANTLR parser generator (\textit{ANTLR})~\cite{}, (c) the minimizing variant of delta debugging (DDMin)~\cite{} and (d) three variants of the maximizing variant of delta debugging (DDMax)~\cite{kirschner2020debugging}.
%%\todo{cite and create command shortcuts for the state-of-the-art}
%
%\noindent \textit{Baseline:}
%
%\noindent \textit{ANTLR:}
%
%\noindent \textit{DDMin:}
%
%\noindent \textit{DDMax:}

\noindent\textbf{Implementation Details:}
We implemented the test infrastructure in about 12k LOC of Java code, \approach was implemented in 765 lines of Java code.
We also slightly modify subject programs %ubject programs needed to be modified slightly 
to provide the required incompleteness feedback (\textit{see} \Cref{tab:subject-programs}).

\noindent\textbf{Platform:} All experiments were conducted
on an ASRock X470D4U with six physical CPU cores and 32GB of RAM, with an AMD Ryzen 5600X @ 3.70GHz, 12~virtual cores, running Debian GNU/Linux.

%\noindent
%\textbf{

\noindent\textbf{Research Protocol:}
We first collect a large corpus of  real-world files which we split into a set of (50) valid files and invalid files. We create additional mutated invalid files by injecting single and up to 16 multiple mutations (random insertions, deletions and byte-flips) into the valid files  
% are then mutated using single and up to 16 multiple mutations (random insertions, deletions and byte-flips), which results in the files 
%shown in 
(\Cref{tab:input-details}).
Then, we run each repair technique on each file and collect the required data, i.e. the run time, number of oracle runs, Levenshtein distance and repair status of each file.
As repair techniques, we employ \emph{Baseline}, %(the built-in error recovery strategy of the subject program), the error recovery strategy of 
\emph{ANTLR}, lexical \ddmax, syntactic \ddmax and \approach.
All experiments were conducted within a timeout of four minutes per input, for each repair technique. % file.
\revise{
The maximum time budget was empirically determined in our preliminary experiments to ensure a balanced evaluation for all techniques. We found that four minutes is a sufficient time budget to evaluate all techniques on most inputs. It is also a reasonable maximum repair time for an end-user. In our experiments, a longer timeout did not result in a significant increase in repairs for all techniques.}

%\subsubsection*{\bf Research Protocol}
%For each input format, we collect real-world invalid input files.
%We also collect 50 valid real-world inputs and mutate those into a set with single and up to 16 multiple mutations.
%Then, we execute all files on the subject programs shown in \Cref{tab:subject-programs}, in order to determine the number of input files which fail for each subject program.
%All subject programs are written in the C programming language and needed to be modified slightly to provide the rich failure feedback required for \approach.
%We proceed to run the repair algorithms on each invalid or mutated input file.
%In particular, we are interested in determining the following:
%%\begin{enumerate}
%%\item[a]
%(1.) \textit{\textbf{Baseline:}} the number of invalid input files which are accepted by a subject program as \textit{valid inputs} (i.e. non-failure-inducing inputs processed by the program without leading to a crash), in order to measure the \textit{effectiveness} of the built-in \textit{error recovery} feature of the program;  and
%%\item[b]
%(2.) \textit{\textbf{ANTLR:}} the number of invalid inputs which are repaired by ANTLR inbuilt \textit{error recovery strategy};
%%\item[c]
%(3) \textit{\textbf{bRepair:}} the number of invalid inputs which are repaired by \brepair.

%All our prototypes are single-threaded.

\section{Experimental Results}
\label{sec:results}
%In this section, we
This section discusses the %findings %
results %and findings 
of our experiments. %mpirical evaluations.
% for each of the posed research questions.

%\todo{RQ1: we need to discuss the number of repair attempts or input fragments tested for feedback for \approach versus ddmax}

\noindent
\textbf{RQ1 Data Recovery and Data Loss:} 
Let us investigate the amount of data recovered and lost by \approach), in comparison to
the best-performing state-of-the-art method (\ddmax).

\noindent
\textbf{\textit{Data Recovery:}} 
%\todo{fix to new evaluation results} \todo{``100.6'' in \autoref{tab:data-recovery} is hard to justify, we need to remove insertions when computing this}	
%For a fair evaluation, % of data recovery, 
We examine 474 inputs that were completely repaired by %all three approaches, i.e., 
lexical \ddmax, syntactic \ddmax and \approach, excluding empty files (and white spaces). \Cref{tab:data-recovery-and-loss} %and \recheck{Figure X
highlights the amount of data %loss and data
recovered and lost % achieved 
by each approach.\footnote{Note that due to input synthesis (i.e., insertion of input elements), \brepair may report recovering more than 100\% of the input file (e.g., 100.6\% for SExp in \Cref{tab:data-recovery-and-loss}).} %, as average over each file. %Our evaluation r

%Evaluation r
Results show that \textit{\approach has a very high data recovery rate}. 
It recovered 91\% of input data, on average. 
This is 23\% more than the most effective baseline (syntactic \ddmax). For all input formats, \approach recovered up to (75\%) more data than both variants of \ddmax. Consider TinyC, where \approach recovered up to 75\% more input data than \ddmax (\textit{see} \Cref{tab:data-recovery-and-loss}). These results show that \approach is more effective in recovering valid input data 
than \ddmax.

\noindent
\textbf{\textit{Data Loss:}} To measure data loss, we compute the Levenshtein distance between repaired and invalid files using %Meanwhile, %For a balanced and tractable evaluation of data loss,
%we employed the
%a set of 
327 completely repaired inputs %by all three approaches which %(321) inputs that
%which the Levenshtein distance implementation could compute %their edit distance
where edit distance could be computed within a threshold of 750 edit distances (about 30 seconds).

We found that \textit{\approach achieves a low data loss of about 11 edit distances, on avarage},  which is \textit{up to 10 times lower than (syntactic) \ddmax}. \Cref{tab:data-recovery-and-loss} shows that 
%this result of this experiment. F
for almost all input formats, \approach achieved a lower data loss than syntactic and lexical \ddmax (except for SExp). This better  performance is attributed to the lightweight failure feedback of \approach, which allows it to distinguish between incomplete and incorrect input fragments.

\begin{result}
\approach has a high data recovery rate (91\%, on average) and its data %The data recovery and
loss is up to 26 times lower than that of ddmax (e.g., INI).
\end{result}

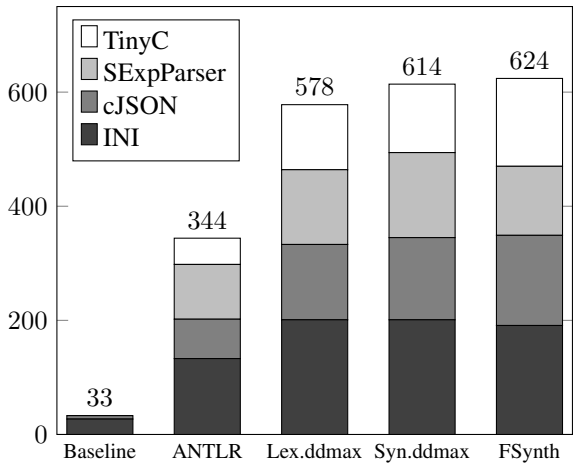
\begin{figure}[!tbp]
%\vspace{-0.1cm}
\centering
\pgfplotstableread{
Label      INI    cJSON    SExpParser   TinyC
Baseline   27     6        0            0
ANTLR      133    69       96           46
Lex.ddmax  201    132      131          114
Syn.ddmax  201    144      149          120
FSynth     191    158      121          154
    }\effectivenessdata
\begin{minipage}{.45\textwidth}
\begin{tikzpicture}
\begin{axis}[
    ybar stacked,
    ymin=0,
    ymax=750,
    xtick=data,
    bar width=25,
    legend style={cells={anchor=west}, legend pos=north west},
    reverse legend=true,
    xticklabels from table={\effectivenessdata}{Label},
    xticklabel style={text width=2cm,align=center,font=\footnotesize},
    xtick style={draw=none},
]
    \addplot [fill=darkgray] table [y=INI, meta=Label, x expr=\coordindex] {\effectivenessdata};
    \addlegendentry{INI}
    \addplot [fill=gray] table [y=cJSON, meta=Label, x expr=\coordindex] {\effectivenessdata};
    \addlegendentry{cJSON}
    \addplot [fill=lightgray] table [y=SExpParser, meta=Label, x expr=\coordindex] {\effectivenessdata};
    \addlegendentry{SExpParser}
    \addplot [fill=white,nodes near coords,point meta=y] table [y=TinyC, meta=Label, x expr=\coordindex] {\effectivenessdata};
    \addlegendentry{TinyC}
\end{axis}
\end{tikzpicture}
\end{minipage}
\caption{Number of files repaired by each approach}
\label{fig:effectiveness}
%\vspace{-0.6cm}
\end{figure}

\noindent
\textbf{RQ2 Effectiveness:} 
%\Cref{tab:input-details} provides details of the %number of
%%real-world and mutated
%invalid inputs used for each input format.
%We also examine the contribution of the insertion  of input elements (i.e., input synthesis) to the effectiveness of our approach.
In this experiment,
%and efficiency of our approach
%Specifically,
we measure the total number of files repaired by our approach. In addition, we compare 
the performance of \approach to 
%Overall, we compare \approach to 
four state-of-the-art methods containing 
%that of %in comparison to
%the number of repairs achieved % compare to %as well as
both \textit{language-agnostic input repair} approaches and \textit{grammar-based input repair} techniques. 
\textit{Language-agnostic input repair} approaches include the built-in error-recovery of the subject program (called \textit{baseline}) and 
%the maximizing variants of delta debugging, i.e., 
\textit{lexical ddmax} (\textit{see}  \Cref{tab:effectiveness-no-grammar}). 
For comparison to \textit{grammar-based input repair methods}, we employ the built-in error-recovery strategy of the \textit{ANTLR} parser generator, and \textit{syntactic ddmax}  (\textit{see}  \Cref{tab:effectiveness-grammar}). 
%four state-of-the-art techniques,
%for all four input formats. We also compare the effectiveness of \approach to that of \recheck{four} state-of-the-art techniques,
%, namely the built-in error-recovery of the subject program (called \textit{baseline}) and \textit{ANTLR}, and % as well as
%the two maximizing variants of delta debugging, i.e., \textit{lexical ddmax} and \textit{syntactic ddmax}.   
\Cref{fig:effectiveness}
%, \Cref{tab:effectiveness-no-grammar} and \Cref{tab:effectiveness-grammar} 
also highlights the effectiveness of \approach in contrast to the state-of-the-art methods. 
%each approach %, as well as their effectiveness
%for each input format.

% to determine the strengths %and complementarity
%of our approach.
% and the time taken to achieve this repair.

%\noindent \textit{Effectiveness:}
%\todo{tables and figures}

\noindent \textbf{\textit{Repair Effectiveness:}} %In our evaluation, %we found that
\textit{Our approach %(\approach) 
is very effective in repairing invalid inputs, it repaired almost four out of every five invalid input withing a four minute timeout}, i.e., % \approach repaired 
about 77\% (624 out of 806) of invalid inputs. In comparison to language-agnostic approaches, %our 
\approach is \textit{up to 18 times more effective than the built-in error recovery strategy of the subject programs (33 vs. 624 repairs)}, and % \approach is 
up to 35\% more effective than the best performing language-agnostic approach for certain input formats (cJSON and TinyC). Overall, 
\Cref{tab:effectiveness-no-grammar} shows that \approach is eight percent more effective than the best language-agnostic repair technique (i.e., lexical \ddmax). 
%Without the knowledge of the input grammar, 
%IN comparison to grammar-based input repair techniques, 
%In our evaluation, w
Additionally, results show that %e found that 
\approach outperforms the %best performing 
grammar-based input repair approaches by up to 28\%. 
% for %. %, for two out of four input formats (
%CJSON and TinyC. 
%). Moreover, i
%It is (20 to 35\%) more effective than language-agnostic techniques for two out of four input formats---CJSON and TinyC.
% input formats. 
%the state-of-the-art 
%Furthermore, 
Despite zero knowledge of the input format, 
\Cref{tab:effectiveness-grammar}
%illustrates these results.
shows that %for most (three out of four) input formats, 
our approach outperformed the grammar-based input repair approaches: \approach is almost twice as effective as the error-recovery strategy of ANTLR (\textit{see} \Cref{fig:effectiveness}), and 
%In addition, 
%Furthermore, %\approach 
%it is 
slightly (2\%) more effective than the best-performing grammar-based approach (i.e., syntactic \ddmax). %We believe tT
The effectiveness of syntactic \ddmax is due to its knowledge of the input grammar. 
%, which reduces the number of required test experiments. 
% its search space for deletion. 
%additional know
%best 
%(except for SExp).  
%\approach is up to 45\% and 29\% more effective than lexical and syntactic ddmax %for two input formats
%(CJSON and TinyC),
%respectively.
% and shows that for
%about 20x as effective as the error recovery of the subject programs (33 repairs), and almost twice as (or 43\% more) effective than ANTLR's inbuilt error-recovery strategy (344 repairs). On the other hand, our approach outperformed the second most effective technique (\textit{ddmax}) by up to \recheck{18\%}. \approach is 18\% and seven percent more effective than lexical ddmax (69\%) and syntactic ddmax (75\%), respectively.
%In summary, t
These results suggest that %(a) \approach is highly effective in repairing invalid input files, (b) our approach more effective than the baselines, and (c)
the \textit{combination of failure feedback and input synthesis} is vital for the effective repair of invalid inputs. 
%, this is evident by the effectiveness of \approach in comparison to the state-of-the-art methods.

\begin{result}
\approach is very effective in repairing invalid input files: %, and more effective than the state-of-the-art.
It repaired four out of five (77\% of) invalid inputs and it is %up to 
8\% more effective than the best language-agnostic method (lexical \ddmax). %baseline.
\end{result}
\done{8\% is not much, and only on average, too. This may not be the best opener, although the later results (specifically data loss) are impressive. -- AZ}

%\begin{table}[!tbp]\centering
%\caption{Details of Repaired Invalid %Failure-inducing
%Inputs}
%\begin{tabular}{|l | l | r | r | r | r | r | l |}
%\hline
%&  \multicolumn{5}{c|}{\textbf{Number of Repaired Inputs}}  \\
%\textbf{Technique(s)} & \textbf{INI} & \textbf{cJSON} & \textbf{SExp} & \textbf{TinyC} & \textbf{Total (\%)} \\
%\hline
%\textbf{Lex. ddmax } & 0 & 3 & 0 & 3 & 6  (0.8\%) \\
%\textbf{Syn. ddmax } & 0 & 6 & 23 & 12 & 41  (5.6\%)  \\
%\textbf{\approach } & 0 & 27 & 0 & 63 & 90 (12.3\%)  \\
%\hline
%\textbf{Lex. $\cap$ Syn.} & 0 & 14 & 3 & 19 & 36  (4.9\%) \\
%\textbf{Lex. $\cap$ Appr.} & 0 & 26 & 0 & 3 & 29  (4.0\%) \\
%\textbf{Syn. $\cap$ Appr.} & 0 & 41 & 1 & 7 & 49  (6.7\%) \\
%\hline
%\textbf{All 3 techs.}
%%Lex. $\cap$ Syn. $\cap$ Appr.}
%& 201 & 79 & 121 & 81 & 482  (65.8\%)\\
%%\textbf{Both} &  201 & 105 & 121 & 84 & 511 (74\%) \\
%%\textbf{Ddmax Only} & 0 & 17 & 3 & 22 & 42 (6\%) \\
%%\textbf{\approach Only} & 0 & 68 & 1 & 70 & 139 (20\%) \\
%\hline
%\textbf{All Repairs} & 201 & 196 & 148 & 188 &  733  (N/A)\\
%\hline
%%\textbf{No Repairs} & 0 & 2 & 11 & 60 & 73 \\
%%\hline
%%\textbf{Total Repaired} & & & & & \\
%%\textbf{Total Repaired} & 201 & 190 &  125 & 176 & 692  \\
%%\hline
%%\textbf{Total Unrepaired} &  0 & 17 & 25 & 72 & 114 \\
%%\hline
%\end{tabular}
%\label{tab:repair-complementarity}
%%\vspace{-0.5cm}
%\end{table}

\begin{table}[!tbp]\centering
\caption{Number of Repaired Inputs within a four minute timeout for \approach vs. Language-agnostic state-of-the-art, i.e., lexical \ddmax and  baseline. The highest number of repaired inputs are in bold, as well as significant percentage improvement (Improv.) of \approach over the best baseline greater than five percent ($>$5\%). }
%\vspace{-0.5cm}
\begin{tabular}{|l | c | r  r  r  r |}
\hline
&  \multicolumn{5}{c|}{\textbf{Number of Repaired Inputs}}  \\
&  \multicolumn{1}{c|}{\textbf{ALL}} & \multicolumn{4}{c|}{\textbf{Input Format}}  \\
\textbf{Techniques} & \textbf{Total} \textbf{(\%)} & \textbf{INI} & \textbf{cJSON} & \textbf{SExp} & \textbf{TinyC} \\
\hline
\textbf{Baseline}   & 33 (4\%) & 27	 & 6 &	0	& 0\\
%\textbf{ANTLR} & 344 (43\%) & 133 & 69 & 96 &  46   \\
\textbf{Lex. ddmax} & 578 (72\%) & \textbf{201}  & 132  & \textbf{131} & 114  \\ 		
\hline	
%\textbf{Syn. ddmax} & \textbf{614} (\textbf{76\%}) & \textbf{201}  & \textbf{144}  & \textbf{149}  & \textbf{120}  \\ 	
%\hline
\textbf{\approach}  & \textbf{624} (\textbf{77\%}) & 191 & \textbf{158}  & 121  & \textbf{154} \\
\hline
\textbf{Improv.} &  \textbf{8\%}  & $-$5\% & \textbf{20\%} & $-$8\% & \textbf{35\%} \\
%\textbf{Improv. vs. Syn.} &  \textbf{2\%}  & -5\% & \textbf{10\%} & -19\% & \textbf{28\%} \\
\hline
%\textbf{Total \#Repairs} & 763  & 510  & 490  & 425 &  2188 \\ 			
%\hline
\end{tabular}
\label{tab:effectiveness-no-grammar}
%\vspace{-0.5cm}
\end{table}

\begin{table}[!tbp]\centering
\caption{Number of Repaired Invalid Inputs within a four minute timeout for \approach vs. Grammar-based input repair approaches, i.e.,  ANTLR and Syntactic \ddmax. The highest number of repaired inputs are in bold, as well as significant percentage improvement (Improv.) of \approach over the best baseline greater than five percent ($>$5\%). }
%\vspace{-0.5cm}
\begin{tabular}{|l | c | r  r  r  r |}
\hline
&  \multicolumn{5}{c|}{\textbf{Number of Repaired Inputs}}  \\
&  \multicolumn{1}{c|}{\textbf{ALL}} & \multicolumn{4}{c|}{\textbf{Input Format}}  \\
\textbf{Techniques} & \textbf{Total} \textbf{(\%)} & \textbf{INI} & \textbf{cJSON} & \textbf{SExp} & \textbf{TinyC} \\
\hline
\textbf{ANTLR} & 344 (43\%) & 133 & 69 & 96 &  46   \\
\textbf{Syn. ddmax} & 614 (76\%) & \textbf{201}  & 144  & \textbf{149}  & 120  \\ 	
\hline
\textbf{\approach}  & \textbf{624} (\textbf{77\%}) & 191 & \textbf{158}  & 121  &  \textbf{154} \\
\hline
\textbf{Improv.} &  2\%  & $-$5\% & \textbf{10\%} & $-$19\% & \textbf{28\%} \\
\hline
\end{tabular}
\label{tab:effectiveness-grammar}
\end{table}

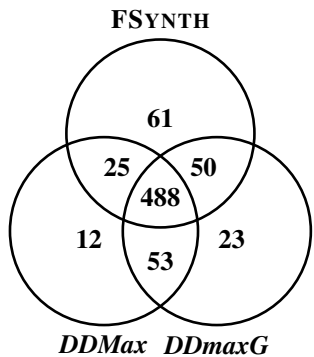
\begin{figure}[!tbp]
%\vspace{-0.4cm}
\centering
\begin{minipage}[b]{0.45\textwidth}
    \centering
    \begin{tikzpicture}[circ/.style={draw=black,line width=1pt,fill=none,shape=circle,minimum width=2.5cm,minimum height=2.5cm},lbl/.style={font=\bfseries}]
        \node[draw=none,minimum width=1.5cm,minimum height=1.29904cm,inner sep=0,outer sep=0] at (0,0) (anchor) {};
        \node[circ] at (anchor.south west) (g1) {};
        \node[circ] at (anchor.south east) (g2) {};
        \node[circ] at (anchor.north) (g3) {};
        \node[lbl,anchor=north] at (g1.south) {\ddmax};
        \node[lbl,anchor=north] at (g2.south) {\ddmaxg};
        \node[lbl,anchor=south] at (g3.north) {\brepair};
        \node[lbl] at ($(g1)!0.5!(g2) - (0,.4)$) {53};
        \node[lbl] at ($(g2)!0.5!(g3) + (.2,.2)$) {50};
        \node[lbl] at ($(g1)!0.5!(g3) + (-.2,.2)$) {25};
        \node[lbl] at ($(anchor.center) - (0,.2)$) {488};
        \node[lbl] at ($(g1.center) - (.2,.1)$) {12};
        \node[lbl] at ($(g2.center) - (-.2,.1)$) {23};
        \node[lbl] at ($(g3.center) + (0,.2)$) {61};
    \end{tikzpicture}
\end{minipage}
\vspace{-0.4cm}
\caption{Venn Diagram showing the number of invalid inputs repaired (solely) by a (combinations of) technique(s) %, or the combination of approaches
}
\label{fig:repair-complementarity}
%\vspace{-0.4cm}
\end{figure}

\noindent \textbf{\textit{Complementarity to \ddmax:}} We further inspect the unique repairs achieved by each approach %, in order 
to understand the complementarity of our approach to the state-of-the-art methods. In this experiment, we inspect the number of unique repairs achieved \textit{solely} by a single approach (e.g., \textit{only} \approach), or two or more approaches (e.g., all approaches). 
%The goal is to determine if there are unique sets of inputs that are solely repaired by a single approach, and not by other approaches. 
\Cref{fig:repair-complementarity} highlights our findings. 

We found that about \textit{one in ten repairs could be solely completed by \approach}, which is three times and five times as many as lexical and syntactic \ddmax %\textit{only} 
repairs, respectively (\textit{see} \Cref{fig:repair-complementarity}).
\approach \textit{solely} completed 9\% (61/712) of all repaired inputs, all of
which \ddmax could not repair. For lexical \ddmax, our approach \textit{solely}
repaired 5$\times$ as many invalid files as lexical \ddmax (61 vs. 12), while
it repairs thrice as many files as syntactic \ddmax (\ddmaxg) (61 vs. 23). Overall, we observe that \textit{\approach complements \ddmax, a combination of these approaches completes 88\% of repairs (712), which is 14\% and 23\% more than the repairs completed \textit{solely} by \approach or syntactic \ddmax, respectively}. 
%These findings 
%show that our approach is complementary to ddmax. 
%It solely completed 61 repairs (9\%), all of which ddmax could not repair. 
%Further inspection 
%suggest that this
\revise{On inspection, we found that the types of repairs completed by \approach (but not by \ddmax) were mostly insertions. We found that the insertions performed by \approach were mostly (missing) syntactic elements of the input like curly braces and colons for \texttt{JSON} or line breaks for \texttt{INI}. In some cases, \approach inserted alphanumeric characters. For instance, if a grammar expects a \texttt{char} where there is a missing \texttt{char}.} 
%Such insertions might alter the data that is contained in the input, for example by adding a value in a JSON dictionary. However, the goal of FSynth (like DDmax) is to provide such candidate repairs for inspection by end-users.
Overall, we attribute the unique repair achieved by \approach to the synergistic combination of \textit{failure feedback} and \textit{input synthesis}. % in \approach.

\begin{result}
\approach %outperforms and 
is complementary to the state-of-the-art methods (lexical and syntactic \ddmax). 
It %is the only technique that can 
solely completes 9\% of all repairs, which is 
up to five times 
%5x 
as much as \ddmax.
\end{result}

%\noindent
%\textbf{RQ3 Diagnostic Quality:} \todo{fix to new evaluation results}
%In this experiment, we evaluate the diagnostic quality of \approach in comparison to the state-of-the-art techniques. In particular, we compare the size of the diagnoses, i.e., the isolated failure cause reported by each approach, as well as the intersection of the diagnoses reported by these approaches. We compare \approach to three main methods, namely \approach, lexical ddmax, syntactic ddmax and ddmin.  \recheck{Table x and figure X} show the diagnostic quality of our approach, in comparison to the state of the art techniques.
%
%We found that
%
%\begin{result}
%XXXX
%XXXX
%\end{result}

\noindent
\textbf{RQ3 Efficiency:} %\todo{fix to new evaluation results}
%\todo{mention that run-time is the execution time of the algorithm, without accounting for the time for experimental pre/post processings }
Let us evaluate the time performance of our approach.
%, i.e., %we investigate
%how long it takes \approach to repair an invalid input, in comparison to the state-of-the-art (ddmax).
%In addition, w
%We also evaluate the number %of repair attempts, i.e., number of
%program runs %test experiments with input fragments,
%performed by each approach.
%To this end, we measure the time performance (i.e., execution time) and the number of program runs performed
%%  of input repairs that were completed
%by \approach, in comparison to lexical ddmax and syntactic ddmax. 
For a balanced evaluation, we
%performed an experiment analysing % where we consider %for
analyse a set of 480 invalid inputs % files
that were completely repaired by all three approaches within %, in less than
the four minutes timeout, %. % (120,000ms).
%Specifically
%To be fair, we report the execution time of each algorithm,
without %accounting for
%the time for
data collection and % or 
experimental
analysis time.
%For program runs, w
%We also compute the number of program runs, i.e., %repair attempts,
%the number of program execution during the test experiments of each approach.
%\recheck{
\Cref{tab:efficiency} %and figure X}
reports the efficiency of all three approaches. % approach compared to ddmax. %the state-of-the-art methods.

%For runtime,

%In this evaluation, w
%We found that 
\textit{\approach is considerably efficient in input repair: It is reasonably fast, it takes about 10 seconds to repair an invalid input, on average}. However, it is about five times slower than \ddmax (two to three seconds, on average). 
%\approach is 50\% more efficient than lexical ddmax (\recheck{1.7s versus 2.5s}), and syntactic ddmax is about three times as efficient as \approach (\recheck{0.6s versus 1.7s}).
%The time performance of
%This implies that \approach is twice as fast as %er than
%lexical ddmax, but twice slower than syntactic ddmax.
%, which is  %)
%the main time-consuming component of \approach, as well as ddmax.
%However,
%\recheck{
\Cref{tab:efficiency} %} %  and Table/Figure X show}
shows %the %Inspecting
%the number of %repair attempts (aka
%program runs performed by each approach. %for all approaches.
%For instance, 
that the execution time
%performance cost is much lower for
of \approach %syntactic ddmax
is much higher %lower
than that of syntactic and lexical \ddmax. 
%and \approach, since
%because \approach requires about three times as many program runs as syntactic ddmax (\recheck{505 vs. 1,586} program runs).
%these approaches is the number of repair attempts and both lexical ddmax and \approach perform more test experiments than syntactic ddmax,
%The lower program runs required by syntactic ddmax is because %since
%it leverages the input grammar to reduce the number of test experiments.
This result is %ese findings are
due to the %can be explained
% is expected since the execution time of %repair attempts (i.e.,
%by the
number of program runs required by \approach, especially due to its additional repair operation (input synthesis) and its extra oracle checks (incompleteness and parser boundary).
%, i.e. test experiments with input fragments.
% %)
%It shows that \approach performed \recheck{X} less repair attempts than lexical ddmax (\recheck{X vs. X} runs), and
%\recheck{X} as many repair attempts as lexical ddmax (\recheck{X vs. X} runs).
%We observe that
%Overall, %these results suggest that
\approach is reasonably efficient, similar to \ddmax, the number of program runs 
%, and %it is
%more efficient than lexical ddmax. %In addition, w
%We observe that the number of program runs is the 
is its main performance bottleneck. %of \approach.  % as well as \ddmax. %all three approaches. % to the
%These results suggest that \approach is %not very time-consuming or
%inexpensive, it as a reasonable number of repair attempts and time performance, especially in  comparison to lexical ddmax.

\begin{result}
%Our
\approach is reasonably fast (10 seconds), but it 
%repairs an invalid input in 10 seconds. 
is 5$\times$ slower than \ddmax because it requires
additional operations and oracle checks. 
%program runs 
%due to input synthesis. 
\end{result}

\begin{table}[!tbp]\centering
\caption{Efficiency of \approach vs. \ddmax, lowest runtime and smaller number of program runs %and significant \% improvement in efficiency
are in \textbf{bold}. %, second-best time performance is in \textit{italics}
}
%\vspace{-0.5cm}
\begin{tabular}{|l |  r  r | r  r |}
\hline
&  \multicolumn{2}{c|}{\textbf{Runtime (s)}} & \multicolumn{2}{c|}{\textbf{\#Prog. Runs}}  \\
\textbf{Techniques} %& \textbf{Inputs}
&  \textbf{Avg.}  & \textbf{Total} & \textbf{Avg.}  & \textbf{Total} \\
\hline
\textbf{Lex. ddmax} & 3.4 & 1653 & 3260 & 1564781 \\
\textbf{Syn. ddmax} & \textbf{2.0} & \textbf{982}  & \textbf{2075}  & \textbf{995802}  \\
\hline
\textbf{\approach} &  10.2 & 4938  & 11537 & 5537601  \\
\hline
\end{tabular}
\label{tab:efficiency}
%\vspace{-0.5cm}
\end{table}

\section{Limitations and Threats to Validity}
\label{sec:threats}

%\todo{discuss threats to validity .... discuss issues like input synthesis (insertion), requires rich feedback: incorrect and incomplete input fragment}

Our approach (\approach) and empirical evaluations may be limited by the following: % validity threats:

\noindent
\textbf{External Validity:} This refers to the \textit{generalizability} of our approach, i.e., 
%results, i.e., 
%We acknowledge 
the threat that \approach may not generalize to other programs and input formats. 
%and their corresponging program written in \<C>. 
%programming languages.
%other than the ones used in this paper. 
To mitigate this threat, we have employed four well-known input formats with varying complexity, their corresponding programs also have  varying sizes (375 to 3062 LOC) and maturity (6 to 21 years old). 

\noindent
\textbf{Internal Validity:} This concerns the \textit{correctness} of our implementation and evaluation, especially if we have correctly implemented \approach and the baselines. We mitigate this threat by testing our implementation of all approaches on small and simple test inputs to ensure the they behave as described. 

\noindent
\textbf{Construct Validity:} This concerns the test oracle and failure feedback employed in our evaluation. To ensure all subjects provide the expected \textit{incomplete} and \textit{(in)correct} feedback, we tested the programs on sample invalid inputs and modified the subject program, if needed. 

\noindent
\textbf{Limited to Data Repair:} The repair produced by our approach aim to ensure maximal data recovery, but it does not ensure that the intended user information is preserved. Hence, \approach is limited to repair of the input data, but not the intended information.

\noindent
\textbf{Input Constraints and Semantics:} \approach does not address concerns about recovering or preserving the input constraints or intended semantics of the input data. For instance, repairing an invalid checksum or hash requires such information, and \approach will be limited for this use cases. However, it would provide repair candidates that allow end-users to debug such semantic issues. 
\revise{
In addition, inputs with significant semantic corruption may not be effectively repaired by FSynth. Even though FSynth is effective in fixing structural parts of invalid inputs, when the corruption is in the semantic part, the missing data becomes difficult to recover. Examples include corrupted numbers in calculations, and dates.}

\noindent
\textbf{Repair via Input Synthesis:} 
\revise{Firstly, the repair via synthesis approach of \approach is exhaustive, thus it can quickly become computationally expensive.} 
Secondly, the repair via synthesis operation %(i.e., insertion) % operation) 
of our approach poses the risk of introducing input elements with unintended semantic consequences. 
%Such operations % into the input, which 
Specifically, insertion operations may lead %cause further 
to data corruption and information distortion. To mitigate this threat, \approach provides several valid candidate repairs ranked by the edit distance 
of each repair candidate from the original input. %to %.
%to the original input. 
Hence, we 
encourage end-users to select the best semantically-fit repair from the 
%employ %\approach  %recommend to 
%the results of %provided by 
% candidates 
%as potential repairs that allows 
%as
potential repair candidates provided by \approach. 

%, out of which to select the best semantically fit repair for their use.
%understanding %and repairing 
%the input data. 

\done{Do you talk about the risk of input synthesis (which may be perceived as greater than the risk of input deletion) somewhere? A small lexical change may have large semantical consequences -- AZ}

\section{Related Work}
\label{sec:related_work}

\noindent
\textbf{Black-box Input Repair:} 
A few techniques have been proposed to analyze input data without program analysis, albeit with the aim of understanding and localizing faults in the program. The earliest works were either focused on simplifying failure-inducing inputs~\cite{zeller2002simplifying, clause2009penumbra}, or isolating fault-revealing input fragments~\cite{hierarchicalDD, sterling2007automated}. Notably, the minimizing delta debugging algorithm (\ddmin) is focused on reducing failure-inducing inputs in order to diagnose and localize faults in the program. More recently, Kirschner et al.~\cite{kirschner2020debugging} proposed a maximizing variant of the delta debugging algorithm (\ddmax) to repair invalid inputs to subsets via deletion, we compare \approach to \ddmax in this work. In contrast to \ddmax, \approach also synthesizes input elements to complete input repair.

\noindent
\textbf{White and Grey-box Input Repair:} Some techniques employ program analysis to fix invalid inputs. As an example, \emph{docovery}~\cite{docovery:ase14} applies symbolic execution to change broken inputs to take error-free paths in the subject program. Similarly, Ammann and Knight~\cite{data_diversity} proposed a method to transform invalid inputs into valid inputs by analyzing the region of the input causing the fault and changing those regions to avoid the fault. Unlike these methods, \approach 
is black-box, it relies on the failure feedback of the subject program.

\noindent
\textbf{Constraint-based Input Repair:} %via Constraint Learning:} 
These approaches automatically learn input constraints then enforce the learned constraints to repair invalid inputs~\cite{hussain2010dynamic, Demsky:2006:IED:1146238.1146266} 
These constraints are often extracted from valid inputs~\cite{Long:2012:AIR:2337223.2337233, Rinard:2007:LCZ:1297027.1297072}, specified with predicates~\cite{elkarablieh2008juzi}, model-based systems~\cite{Demsky:2003:ADR:949343.949314}, goal-directed reasoning~\cite{1553560}, dynamic symbolic execution~\cite{hussain2010dynamic} or invariants~\cite{Demsky:2006:IED:1146238.1146266}. For instance, \emph{S-DAGs}~\cite{scheffczyk2004s} enforce constraints on invalid inputs in a semi-automatic way. Unlike these approaches, \approach does not learn input constraints, it employs input synthesis and failure feedback to fix invalid inputs. 

\noindent
\textbf{Parser-directed Input Repair:}
This refers to the input repair schemes of parsers, interpreters and compilers~\cite{parr2011ll, diekmann2020dont, aho1972minimum, hammond1984survey, backhouse1979syntax}. 
%In most of the syntactic recovery approaches, various different 
These techniques employ operations such as insertion, deletion and replacement of symbols~\cite{anderson1981locally, cerecke2003locally, anderson1983assessment}, extending forward or backwards from a parser error~\cite{burke1982practical, mauney1982forward}, or more general methods of recovery and diagnosis~\cite{krawczyk1980error, aho1972minimum}. 
In this work, we compare \approach to the recovery scheme of the ANTLR parser generator~\cite{parr2011ll} which leverages the input grammar to guide input repair. %Compared to \approach, t
Unlike \approach which aims to fix an invalid input, these schemes need an input grammar, and aim to ensure the compiler does not halt while parsing. 

%\noindent
%\textbf{Data Analysis:} This is the process of analyzing (complex) database systems to remove noisy data, fill in missing data~\cite{xiong2006enhancing, hernandez1995merge}, or identify program errors caused by well-formed but incorrect data~\cite{mucslu2013data}. This challenge is often addressed %manually by allowing developers write and 
%applying logical rules on the database~\cite{pochampally2014fusing, galhardas2000ajax, golab2010data, jeffery2006pipelined, raman2001potter, luebbers2003systematic}. For instance, in continuous data testing (CDT)~\cite{mucslu2015preventing} warn users of test failures likely data errors by executing domain-specific test queries. % to .
%Likewise, \emph{DATAXRAY}~\cite{wang2015error} %also investigates the underlying conditions that cause data bugs, it0
%reveals hidden connections and common properties among data errors. In contrast to \approach which repairs raw user inputs, these approaches aim to %guard data from new errors by 
%detect or repair %ing 
%data errors in database systems. % during modification or repair database systems.
%%
%\noindent
%\textbf{XX}
%

\section{Conclusion}
\label{sec:conclusion}
%\recheck{
%In this paper, we 
This paper presents \approach, an input repair approach that employs input synthesis and lightweight failure feedback (i.e., incomplete and (in)correct checks) to repair invalid inputs. Our approach is black-box, does not require program analysis or an input grammar. We evaluate the data recovery,effectiveness, and efficiency of our approach, in comparison to the state-of-the-art methods. We show that \approach has a very high data recovery rate, it recovered 91\% of input data. 
It is also very effective and efficient in input repair---it completes the repair of about four in five invalid inputs (77\%) within four minutes. Furthermore, we demonstrate that our approach is up to 35\% more effective than the best 
%In comparison to the 
best baseline---(syntactic) \ddmax, without using an input grammar. In summary, this work demonstrates that combining lightweight failure feedback and input synthesis are important for the effective repair of invalid inputs, especially in the absence of an input specification 
%(e.g., grammar)
and program analysis.

In the future, we plan to investigate how to improve the performance of our approach by learning input semantics and constraints.
%, or via program analysis and the knowledge of the input specification. %grammar.
%}
%  more effective thja
%\todo{discuss potential future works}
%\revise{
We also provide our implementation, data and experimental results for easy replication, scrutiny and reuse:
%}
%\todo{we should provide a github page, including an anonymous notebook that allows to test the examples (in Table 1), it will be interesting}

 \begin{center}
% \vspace{-0.2mm}
     \textbf{\url{https://github.com/vrthra/fsynth-artifact}}%\todo{update link}
 \end{center}

% \begin{acks}
% % To Robert, for the bagels and explaining CMYK and color spaces.
% \end{acks}

%\newpage

\balance
\bibliographystyle{IEEEtran}
\bibliography{icse23-brepair}

\end{document}